\documentclass[manuscript]{aastex}

\shorttitle{Two-step Emergence of the Magnetic Flux Sheet}
\shortauthors{Toriumi \& Yokoyama}

\usepackage{subfigure}

\begin{document}


\title{Two-step Emergence of the Magnetic Flux Sheet\\
 from the Solar Convection Zone}


\author{S. Toriumi and T. Yokoyama}
\affil{Department of Earth and Planetary Science, University of Tokyo,
Bunkyo-ku, Tokyo 113-0033, Japan}
\email{toriumi@eps.s.u-tokyo.ac.jp}




\begin{abstract}
We perform two-dimensional MHD simulations of the flux emergence
from the solar convection zone to the corona.
The flux sheet is initially located moderately deep
in the adiabatically stratified convection zone
($-20,000\ {\rm km}$)
and is perturbed to trigger the Parker instability.
The flux rises through the solar interior
due to the magnetic buoyancy,
but suffers a gradual deceleration and a flattening
in the middle of the way to the surface
since the plasma piled on the emerging loop cannot pass through
the convectively stable photosphere.
As the magnetic pressure gradient enhances,
the flux becomes locally unstable to the Parker instability
so that the further evolution to the corona occurs.
The second-step nonlinear emergence is well described
by the expansion law by \citet{shiba89}.
To investigate the condition for this `two-step emergence' model,
we vary the initial field strength and the total flux.
When the initial field is too strong,
the flux exhibits the emergence to the corona
without a deceleration at the surface
and reveals an unrealistically strong flux density
at each footpoint of the coronal loop,
while the flux fragments within the convection zone
or cannot pass through the surface
when the initial field is too weak.
The condition for the `two-step emergence'
is found to be $10^{21}-10^{22}\ {\rm Mx}$
with $10^{4}\ {\rm G}$ at $z=-20,000\ {\rm km}$.
We have some discussions in connection
with the recent observations
and the results of the thin-flux-tube model.
\end{abstract}





\section{Introduction\label{intro}}


Solar active regions are generally
thought to be the consequence of the flux emergence,
i.e., dynamo-generated magnetic fluxes risen from the
deep convection zone
due to the magnetic buoyancy force \citep{par55}.
Observations indicate that
the flux tube should be in a coherent form
and strong enough so as not to be disintegrated
by the turbulent motion during its emergence
through the convection zone.
For generating such a strong flux,
the sub-adiabatically stratified overshoot region
at the bottom of the convection zone has been
suggested to be the suitable place \citep{par75}.
Therefore, a flux emergence should be understood
as a whole process
from the base of the convective layer
to the upper atmosphere
through the surface.

In last two decades, various series of numerical experiments
have been carried out to investigate
the physics of the flux emergence.
For the local evolution above the surface,
the pioneering work was done by \citet{shiba89},
who performed two-dimensional (2D)
magnetohydrodynamic (MHD) simulations of flux emergence
through the undular mode of magnetic buoyancy instability
($\mbox{\boldmath $k$}\parallel \mbox{\boldmath $B$}$,
where $\mbox{\boldmath $k$}$ and $\mbox{\boldmath $B$}$ denote
the wavenumber and the initial magnetic field vector, respectively;
the Parker instability)
to reproduce some dynamical features
such as the rise motion of an arch filament system
and downflows along the magnetic field lines.
Since then, the evolution of an emerging flux
and the interaction with pre-existing coronal fields
have been studied by 2D and 3D simulations
\citep{noza92,matsu93,yoko95,yoko96,fan01b,che08}.
\citet{noza92} performed the emergence from the convectively
unstable solar interior (the convective-Parker instability),
while \citet{yoko95,yoko96} studied the reconnection
between the expanding loop and the pre-existing fields
in the corona and the subsequent formation of X-ray jets.
Three-dimensional calculations by \citet{matsu93} were
performed for the studies of
the interchange ($\mbox{\boldmath $k$}\perp \mbox{\boldmath $B$}$) and
quasi-interchange ($\mbox{\boldmath $k$}\parallel\mbox{\boldmath $B$}$
with $kH_{\rm ph}\ll 1$, where $H_{\rm ph}$ is the photospheric pressure
scale height) mode instabilities.
\citet{fan01b} compared her 3D simulation results
with observed features of newly emerged active regions.
\citet{che08} found that the numerical modeling of emerging flux regions
by 3D radiative MHD simulations exhibits photospheric characteristics
that are comparable with the observations
from the {\it Hinode}/Solar Optical Telescope (SOT).
These simulations assumed that
the initial flux is embedded
just below the photosphere ($\ga -2000\ {\rm km}$)
a priori as a horizontal sheet or a twisted tube.
It is because their experiments mainly aimed
to clarify the local behaviors in the solar atmosphere.
Such an initial structure, however, is not obvious
since the initial form depends on its history
of pushing through the convection zone.
Therefore, a numerical experiment
including the evolution from a substantial depth
has been needed.

The simulations focusing on the flux emergence
within the relatively deep solar interior
have been done by using the thin-flux-tube approximation
\citep{spr81,dsil93,cal95}.
One of the most important conclusions obtained from the various
thin-flux-tube simulations is that rising tubes with small magnetic flux
(below 10$^{21}$ Mx for 10$^{4}$ G at the base) cannot reach the
photosphere because the apices of the loops loose magnetic fields and
subsequently `explode' \citep{mor95}.
By assuming the anelastic approximation \citep{gou69,lan99},
\citet{fan01a} computed the evolution of the 3D undulatory instability
of the horizontal magnetic flux layer,
which formed arch-like magnetic tubes
with downflows from their apices to the troughs.
Note that both types of approximations
(thin-flux-tube and anelastic)
are not applicable in the upper convection zone
($\ga -30\ {\rm Mm}$)
where the diameter of the flux tube exceeds
the local pressure scale height
and the flow velocity becomes close to the sound speed.

\citet{abb03} calculated a flux emergence
by connecting the anelastic MHD convective layer
and the fully compressible MHD solar atmosphere
from the photosphere to the low corona.
However, their full MHD atmosphere
did not include the upper convection zone.
To investigate the detailed behavior
of the emerging flux
from the convection zone to the atmosphere,
we have to deal with the full MHD numerical box
including the convective layer,
the photosphere/chromosphere, and the corona.

The dynamical behavior of an emerging flux
including both the interior and the upper atmosphere
is not yet clear,
but is thought to obey the following picture,
which we name `two-step emergence' model
\citep[cf.][]{matsu93,maga01}.
Magnetic fluxes emerged from the bottom of the convection zone are
depressed and decelerated by the sub-adiabatic photosphere
and extended horizontally around the photosphere/chromosphere.
Meanwhile, fluxes are still transported from below to enhance the magnetic pressure gradient.
Finally, the fluxes above the photosphere become unstable to the magnetic buoyancy instability so that the further evolution into the corona occurs.
By the development of supercomputers,
full MHD simulations in the convection zone without approximations
have come to be realizable.
The purpose of this work is to investigate the `two-step emergence' model numerically.

Several experiments have confirmed this `two-step' model.
\citet{maga01} studied the emergence of the magnetic flux tube from the convection zone by means of 2.5-dimensional MHD simulations focused on the cross section of the tube.
He found the deceleration of the rising flux tube due to the
convectively stable photosphere and the subsequent horizontal outflow.
\citet{arc04} performed 3D simulations using the
criterion by \citet{ach79} to analyze the magnetic
buoyancy instability within the photosphere/chromosphere, while \citet{mur06} did parameter studies of the dependence of the initial magnetic field strength of the tube and its twist, finding that the tube evolves in the self-similar way when varying the field strength and that the magnetic buoyancy instability and the second-step evolution do not occur when the field is too weak.
In these studies, the initial flux tubes are located
in the uppermost convective region at some 1000 km depth.
Recently, Solar Optical Telescope (SOT) aboard the {\it Hinode} satellite
\cite[e.g.][]{tsune08} observed small-scale magnetic flux emergence.
Thanks to high-resolution and high-cadence multi-wavelength
observations, \citet{otsu10} found the deceleration
of the apex of the arch filament system in the chromosphere,
which can be the possible evidence of the two-step model.

In this study,
we perform two-dimensional fully compressible MHD simulations
to investigate the dynamical evolution
from the adiabatically stratified convective layer
into the corona through the isothermal
(strongly sub-adiabatic, i.e., convectively stable)
photosphere/chromosphere.
We set the initial magnetic flux sheet
in the moderately deep convection zone at 20 Mm depth,
not at the bottom of the solar interior,
because the emergence from the base is beyond
the computation ability.
The numerical results reproduce the two-step model well.
However, the picture of the two-step emergence is
much far from the previous studies
\citep[e.g.][]{maga01,arc04,mur06}.
The location of the initial flux is so deep
that the typical wavelengths of the Parker instability
($10-20$ times the local pressure scale height)
are different between the primary and the secondary emergence;
the first-step evolution occurs at $z=-20\ {\rm Mm}$
and its typical wavelength is about $100\ {\rm Mm}$,
while, at the photosphere, the second-step emergence
has its wavelength of the order of a few $1000\ {\rm km}$.
We also discuss the dependence of the flux sheet's behavior
(`two-step emergence', `direct emergence' or `failed emergence') on its
magnetic field strength and amount of fluxes
through the parameter survey.

The numerical setup and the assumed conditions used in this study
are presented in Section \ref{model}.
We show the typical case of the `two-step emergence'
in Section \ref{general},
while Section \ref{parameter} gives the results of the parameter survey
and make some discussions with preceding studies
and recent observations.
Finally, in Section \ref{conclude}, we summarize the study.

\section{Numerical Model\label{model}}

\subsection{Assumptions and Basic Equations}


We consider an isolated magnetic flux sheet located in a
convectively marginally stable gas layer in a two-dimensional
Cartesian coordinate system ($x,\,z$),
where $z$-direction is antiparallel to the
gravitational acceleration.
We solve adiabatic two-dimensional
($\partial/\partial y=0,\,B_{y}=0,\,V_{y}=0$)
magnetohydrodynamic (MHD) equations.
The basic equations are as follows;
\begin{eqnarray}
	\frac{\partial\rho}{\partial t}
	 + \mbox{\boldmath $\nabla$}
	 \cdot(\rho \mbox{\boldmath $V$})=0\ ,
\end{eqnarray}
\begin{eqnarray}
	\frac{\partial(\rho \mbox{\boldmath $V$})}{\partial t}
	+ \mbox{\boldmath $\nabla$}\cdot
	\left(
	 \rho \mbox{\boldmath $VV$}+p\mbox{\boldmath $I$} 
	 -\frac{\mbox{\boldmath $BB$}}{4\pi}
	 +\frac{\mbox{\boldmath $B$}^{2}}{8\pi}\mbox{\boldmath $I$}
	\right)
	-\rho \mbox{\boldmath $g$} =0\ ,
\end{eqnarray}
\begin{eqnarray}
	\frac{\partial \mbox{\boldmath $B$}}{\partial t}
	= \mbox{\boldmath $\nabla$}\times
	(\mbox{\boldmath $V$}\times \mbox{\boldmath $B$})\ ,
\end{eqnarray}
\begin{eqnarray}
	\frac{\partial}{\partial t}
	\left(
	 \rho U+\frac{1}{2}\rho \mbox{\boldmath $V$}^{2}
	 +\frac{\mbox{\boldmath $B$}^{2}}{8\pi}
	\right)
	+\mbox{\boldmath $\nabla$}\cdot
	\left[
	 (\rho U+p+\frac{1}{2}\rho \mbox{\boldmath $V$}^{2})
	 \mbox{\boldmath $V$}
	 +\frac{c}{4\pi}\mbox{\boldmath $E$}\times \mbox{\boldmath $B$}
	\right]
	-\rho \mbox{\boldmath $g$}\cdot \mbox{\boldmath $V$}=0\ ,
\end{eqnarray}
and
\begin{eqnarray}
	U=\frac{1}{\gamma -1}\frac{p}{\rho}\ ,
\end{eqnarray}
\begin{eqnarray}
	\mbox{\boldmath $E$}
	 =-\frac{1}{c}\mbox{\boldmath $V$}\times \mbox{\boldmath $B$}\ ,
\end{eqnarray}
where $U$ is the internal energy per unit mass,
$\mbox{\boldmath $I$}$ is the unit tensor,
$\mbox{\boldmath $g$}=(0,0,-g_{0})$
($g_{0}$ is constant and its value is given below)
is the gravitational acceleration,
and other symbols have their usual meanings.
We assume a ratio of specific heats, $\gamma$, of 5/3.

\subsection{Initial Conditions}

\begin{figure}
\begin{center}
\subfigure[]{\epsscale{0.8}\plotone{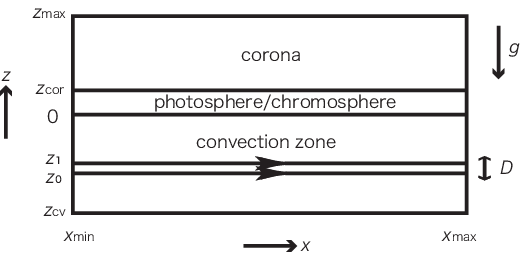}
\label{fig:ini-a}}\\
\subfigure[]{\epsscale{0.8}\plotone{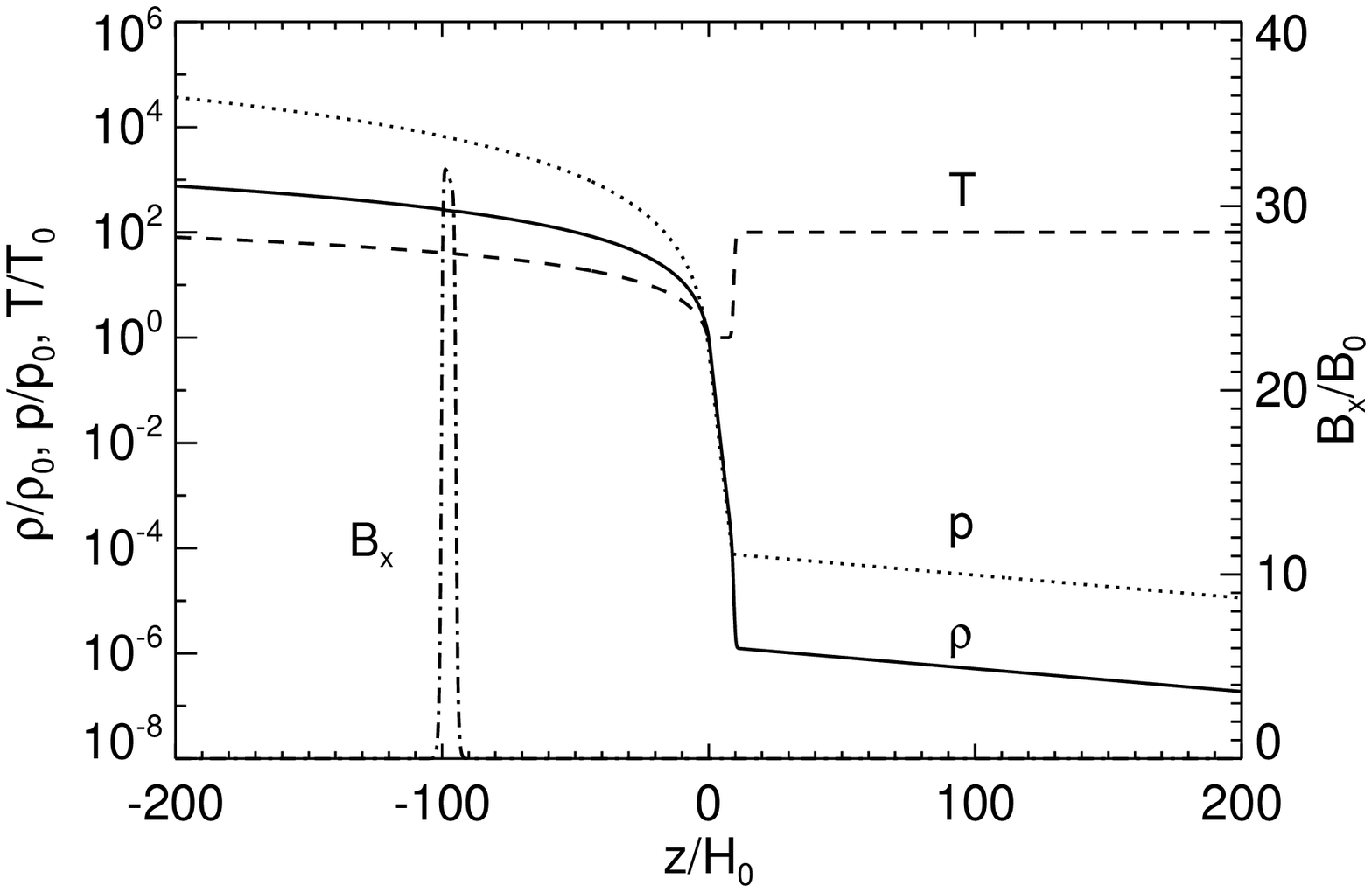}
\label{fig:ini-b}}
\caption{\subref{fig:ini-a} Schematic depiction of the initial set-up.
 \subref{fig:ini-b} One-dimensional ($z$-)distributions of the initial density
 (solid line), pressure (dotted line), temperature (dashed line),
 and magnetic field strength (dashed-dotted line).}
\label{fig:ini}
\end{center}
\end{figure}

The initial magnetostatic gas layer is composed of
three regions:
hot and cold isothermal layers
in upper and middle regions represent
a very simplified model of the solar corona
and photosphere/chromosphere,
and a non-isothermal layer in the lower region
models the convection zone (see Fig. \ref{fig:ini-a}).
We take $z=0$ to be the base height of the photosphere.
The units of length, velocity, time, and density
in the simulations
are $H_{0}$, $C_{s0}$, $H_{0}/C_{s0}\equiv \tau_{0}$,
and $\rho_{0}$, respectively,
where $H_{0}=k_{\rm B}T_{0}/(mg_{0})$ is
the pressure scale height
($k_{\rm B}$ is Boltzmann constant
and $m$ is mean molecular mass),
$C_{s0}$ the sound speed,
and $\rho_{0}$ the density at $z=0$ in the photosphere.
The gas pressure, temperature, and magnetic field strength
are normalized by the combinations of the units above,
i.e., $p_{0}=\rho_{0}C_{s0}^{2}$,
$T_{0}=mC_{s0}/(\gamma k_{\rm B})$,
and $B_{0}=(\rho_{0}C_{s0}^{2})^{1/2}$, respectively.
The gravity is given as $g_{0}=C_{s0}^{2}/(\gamma H_{0})$
by definition.
For the comparison of numerical results with observations,
we use $H_{0}=200\ {\rm km}$, $C_{s0}=8\ {\rm km\,s}^{-1}$,
$\tau_{0}=H_{0}/C_{s0}=25\ {\rm s}$,
and $\rho_{0}=1.4\times 10^{-7}\ {\rm g\ cm}^{-3}$,
which are typical values
for the solar photosphere and chromosphere.
Then, $p_{0}=9.0\times 10^{4}\ {\rm dyn\ cm}^{-2}$,
$T_{0}=4000\ {\rm K}$, and $B_{0}=300\,{\rm G}$.

The initial temperature distribution of
the photosphere/chromosphere and the corona $(z\geq 0)$
is assumed to be
\begin{equation}
 T(z)=T_{\rm ph}+(T_{\rm cor}-T_{\rm ph})
  \left\{
   \tanh{[(z-z_{\rm cor})/w_{\rm tr}]+1}
  \right\}/2\ ,
\end{equation}
where $T_{\rm cor}$ and $T_{\rm ph}$ are the
respective temperatures in the corona and
in the photosphere/chromosphere,
$z_{\rm cor}$ is the height of the base of the
corona, and $w_{\rm tr}$ is the temperature scale
height of the transition region.
We take $T_{\rm cor}=100T_{0}$, $T_{\rm ph}=T_{0}$,
$z_{\rm cor}=10H_{0}$, and $w_{\rm tr}=0.5H_{0}$.
As for the initial temperature distribution
in the convective layer ($z\leq 0$),
we assume
\begin{equation}
 T(z)=T_{\rm ph}-\alpha z
  \left|
   \frac{dT}{dz}
  \right|_{\rm ad}\ ,
\end{equation}
where
\begin{equation}
  \left|
   \frac{dT}{dz}
  \right|_{\rm ad}
  =\frac{\gamma -1}{\gamma}\frac{mg_{0}}{k_{\rm B}}
\end{equation}
is the adiabatic temperature gradient,
and $\alpha$ is a parameter of
the temperature gradient of the convection zone.
In our calculations, $\alpha$ is set to be unity,
i.e., the initial temperature distribution
in the convection zone is adiabatic.

The magnetic field is initially horizontal,
$\mbox{\boldmath $B$}=(B_{x}(z),0,0)$,
and is embedded in the convection zone.
The distribution of magnetic field strength is
given by
\begin{equation}
 B_{x}(z)=[8\pi p(z)/\beta(z)]^{1/2}\ ,
\end{equation}
where
\begin{equation}
 \beta(z)=\beta_{\ast}/f(z)\ ,
\end{equation}
and
\begin{equation}
 f(z)=\frac{1}{4}
  \left[
   \tanh{\left(
	  \frac{z-z_{0}}{w_{0}}
	 \right)}+1
  \right]
  \left[
   -\tanh{\left(
	   \frac{z-z_{1}}{w_{1}}
	  \right)}+1
  \right]\ .
  \label{eq:f}
\end{equation}
Here $\beta_{\ast}$ is the ratio of gas pressure
to magnetic pressure at the center of the magnetic flux sheet,
and $z_{0}$ and $z_{1}=z_{0}+D$ are the heights of the
lower and upper boundaries of the flux sheet,
where $D$ is the vertical thickness of the sheet.
We use $z_{0}=-100H_{0}\simeq -20,000\,{\rm km}$.
In all of our calculations,
$w_{0}$ and $w_{1}$ are set to be $0.5H_{0}$.
We take $\beta_{\ast}=160$ and
$D=5H_{0}\simeq 1000\,{\rm km}$ for case 1 (the typical model),
so as the initial magnetic field
strength $B_{x}$ to be $10^{4}\,{\rm G}$ and
the total magnetic flux $\Phi$ to be $10^{21}\,{\rm Mx}$.
To calculate the total magnetic flux,
we regard the initial flux sheet as a rectangular
prism with a base $D$ (in $z$ direction)
by $10D$ (in $y$ direction).

On the basis of the initial plasma $\beta$
distribution mentioned above,
the initial density and pressure distribution
are calculated numerically by using
the equation of static pressure balance:
\begin{equation}
 \frac{d}{dz}
  \left[
   p+\frac{B_{x}^{2}}{8\pi}
  \right]+\rho g_{0}=0\ .
\end{equation}
The initial temperature, density, pressure,
and magnetic field strength distributions
for the typical case are shown in Fig. \ref{fig:ini-b}.

In order to trigger the Parker instability \cite[]{par66}
in the convection zone,
small density perturbations of the form
\begin{equation}
 \delta\rho=Af(z)\rho(x,z)\cos{(2\pi x/\lambda)}
\end{equation}
are initially reduced from the magnetic flux sheet
($z_{0}\leq z\leq z_{1}$)
within the finite horizontal domain
($-3\lambda/4<x<3\lambda/4$),
where $\lambda(=400H_{0})$ is the perturbation wavelength,
and $A(=0.01)$ is the maximum value of the initial
density reduction.
The definition of $f(z)$ is given in equation (\ref{eq:f}).

\subsection{Boundary Conditions and Numerical Procedures}


The domain of the simulation box is
$(x_{\rm min}<x<x_{\rm max})$
and $(z_{\rm min}<z<z_{\rm max})$,
where $x_{\rm min}=-400H_{0}$, $x_{\rm max}=400H_{0}$,
$z_{\rm min}=-200H_{0}$, and $z_{\rm max}=200H_{0}$,
i.e., the total size of the box is
$160\ {\rm Mm}\times 80\ {\rm Mm}$.
This is much larger than those of
the calculations focusing on the emergence
from the uppermost convection zone
to the corona \cite[e.g.][etc.]{shiba89}.
Periodic boundaries are assumed
for $x=x_{\rm min}$ and $x=x_{\rm max}$,
symmetric boundaries for $z=z_{\rm min}$ and $z=z_{\rm max}$.
A damping zone is attached near the top boundary
to reduce the effects of reflected waves.

To solve the equations numerically,
we use the modified Lax-Wendroff scheme.
The code is a part of the numerical package CANS
(Coordinated Astronomical Numerical Software)
maintained by Yokoyama et al
\footnote{
CANS (Coordinated Astronomical Numerical Software)
is available online at:
\url{http://www-space.eps.s.u-tokyo.ac.jp/$\sim$yokoyama/etc/cans/}
}
.
For the typical model (case 1),
the total number of grid points is
$(N_{x}\times N_{z})=(1536\times 1920)$,
and the mesh sizes are
$\Delta x=0.52H_{0}$ and $\Delta z=0.21H_{0}$,
both of which are uniform.
For comparison, we perform other simulations
with different values of the parameters.
We vary the values of the initial magnetic field
strength $B_{x}$ and of the total magnetic flux $\Phi$
of the emerging flux sheet
by adjusting $\beta_{\ast}$ and $D$.
In these calculations
we set the total number of grid points
$(N_{x}\times N_{z})=(1024\times 1280)$,
and the uniform mesh sizes
$\Delta x=0.78H_{0}$ and $\Delta z=0.31H_{0}$.
The cases we examine are summarized in Table \ref{tab:param}.

\section{General Evolution\label{general}}

\begin{figure}
\begin{center}
\subfigure{\includegraphics[clip,scale=0.5]{f2a.eps2}
\label{fig:ro-a}}~
\subfigure{\includegraphics[clip,scale=0.5]{f2b.eps2}
\label{fig:ro-b}}\\
\subfigure{\includegraphics[clip,scale=0.5]{f2c.eps2}
\label{fig:ro-c}}~
\subfigure{\includegraphics[clip,scale=0.5]{f2d.eps2}
\label{fig:ro-d}}\\
\subfigure{\includegraphics[clip,scale=0.5]{f2e.eps2}
\label{fig:ro-e}}~
\subfigure{\includegraphics[clip,scale=0.5]{f2f.eps2}
\label{fig:ro-f}}\\
\subfigure{\includegraphics[clip,scale=0.5]{f2g.eps2}
\label{fig:ro-g}}~
\subfigure{\includegraphics[clip,scale=0.5]{f2h.eps2}
\label{fig:ro-h}}\\
\includegraphics[clip,scale=0.6,angle=-90.]{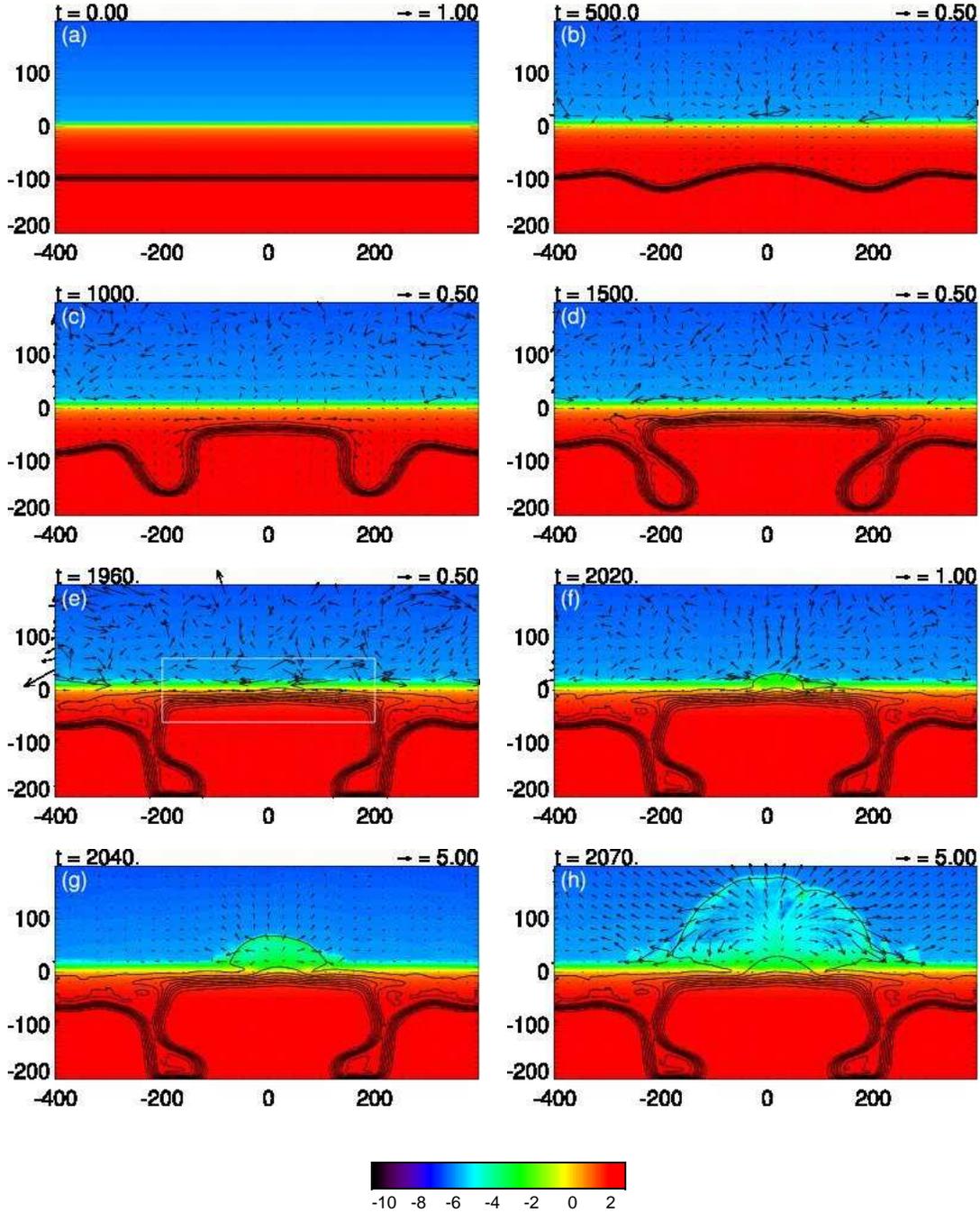}
\caption{Time-evolution of the `two-step emergence' for case 1
(typical model).
\subref{fig:ro-a} $t/\tau_{0}=0$;
\subref{fig:ro-b} $t/\tau_{0}=500$;
\subref{fig:ro-c} $t/\tau_{0}=1000$;
\subref{fig:ro-d} $t/\tau_{0}=1500$;
\subref{fig:ro-e} $t/\tau_{0}=1960$;
\subref{fig:ro-f} $t/\tau_{0}=2020$;
\subref{fig:ro-g} $t/\tau_{0}=2040$;
\subref{fig:ro-h} $t/\tau_{0}=2070$.
Logarithmic density profiles ($\log_{10}{(\rho/\rho_{0})}$)
are indicated by color contour,
while magnetic field lines and velocity vectors are overplotted with
black lines and arrows.
The white box at $t/\tau_{0}=1960$ shows the area we analyze the Parker
 instability (see Fig. \ref{fig:newcomb}).
This figure is also available as an avi animation in the electronic
edition of the {\it Astrophysical Journal}.}
\label{fig:ro}
\end{center}
\end{figure}

\begin{figure}
\epsscale{1.}
\plotone{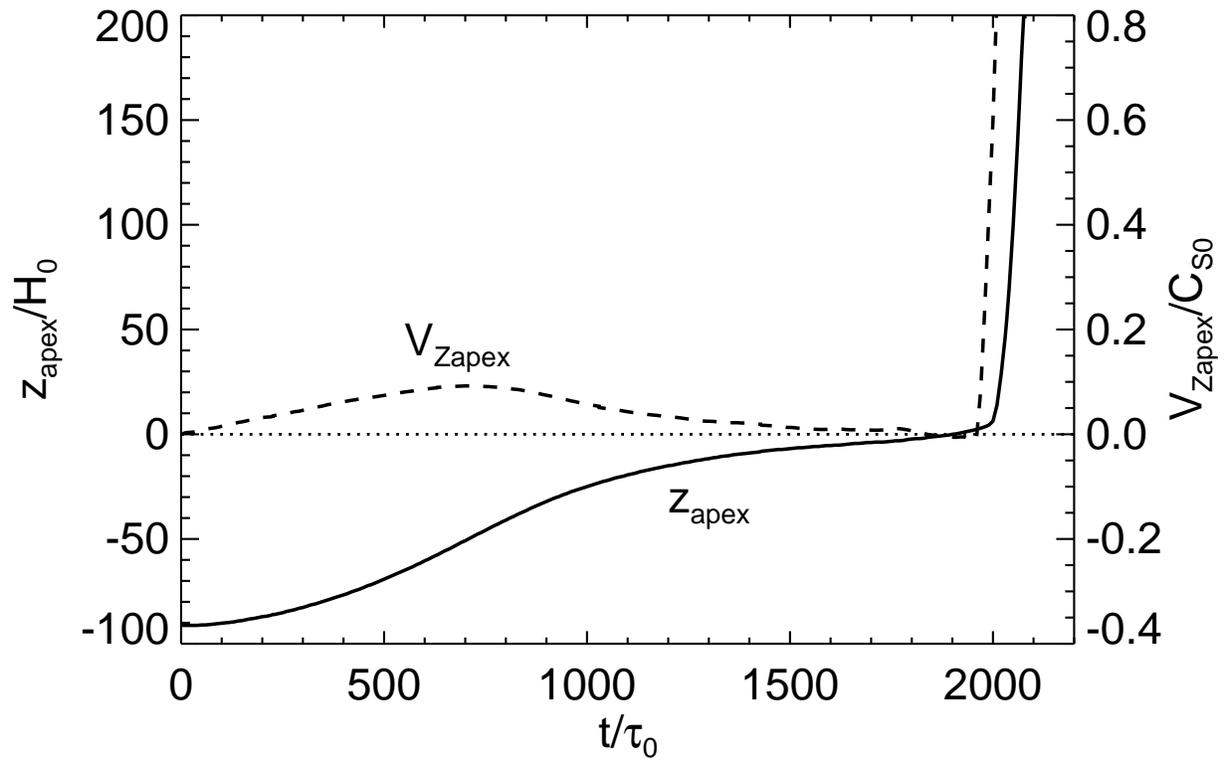}
\caption{Time evolution of the apex of the flux sheet; solid
line is a height of the apex, whose rise velocity is indicated
with dashed line. Photospheric level ($z/H_{0}=0$) is
overplotted with dotted line.}
\label{fig:z_vz_top}
\end{figure}

In this section we show the numerical results of the typical case
(case 1: $B_{x}=10^{4}{\rm\,G}$ and $\Phi=10^{21}{\rm\,Mx}$);
the results display the `two-step emergence'.
Figure \ref{fig:ro} illustrates the development
of the density profile with magnetic field lines and velocity vectors,
while Figure \ref{fig:z_vz_top} indicates
the height of the apex of the magnetic field $z_{\rm apex}$
and its rise velocity $V_{z{\rm apex}}$.
The time evolution can be divided into four stages
according to the rise velocity at the apex of the emerging field.
In the first stage ($0<t/\tau_{0}<700$),
the magnetic flux begins an emergence in the convection zone
due to the Parker instability driven by the magnetic buoyancy
(Fig. \ref{fig:ro-b}).
The rise velocity increases continuously in this stage
(Fig. \ref{fig:z_vz_top}).
The second stage ($700<t/\tau_{0}<1900$) is characterized
by a gradual deceleration (Fig. \ref{fig:z_vz_top})
because the arch-like emerging field becomes deformed to horizontal
so that the mass on the apex
can no longer fall down along the field lines,
and thus continues to pile up on the horizontal field.
As a consequence, the emerging flux stretches around the solar surface
(Figs. \ref{fig:ro-c} and \subref{fig:ro-d}). 
In the third stage ($1900<t/\tau_{0}<2000$),
the top part of the emerging magnetic flux almost stops at the surface
while fluxes are still emerging from below,
that is, the magnetic pressure gradient on
the upper boundary of the flux sheet continues to enhance.
When the magnetic pressure gradient gets steepened enough,
the Parker instability sets in and
drives the further evolution into the upper atmosphere
(Figs. \ref{fig:ro-e} and \subref{fig:ro-f}).
In the final, fourth stage ($t/\tau_{0}>2000$),
the magnetic flux evolves to the corona
due to magnetic pressure,
which is consistent with the results of
classical calculations \citep[e.g.][]{shiba89}
(Figs. \ref{fig:ro-g} and \subref{fig:ro-h}).
In the following, we will discuss each stage in more detail, and examine
the dynamical structure of the expanding loop and the related forces
acting on the magnetic field.

\subsection{First Stage ($0<t/\tau_{0}<700$)}
\begin{figure}
\epsscale{1.}
\plotone{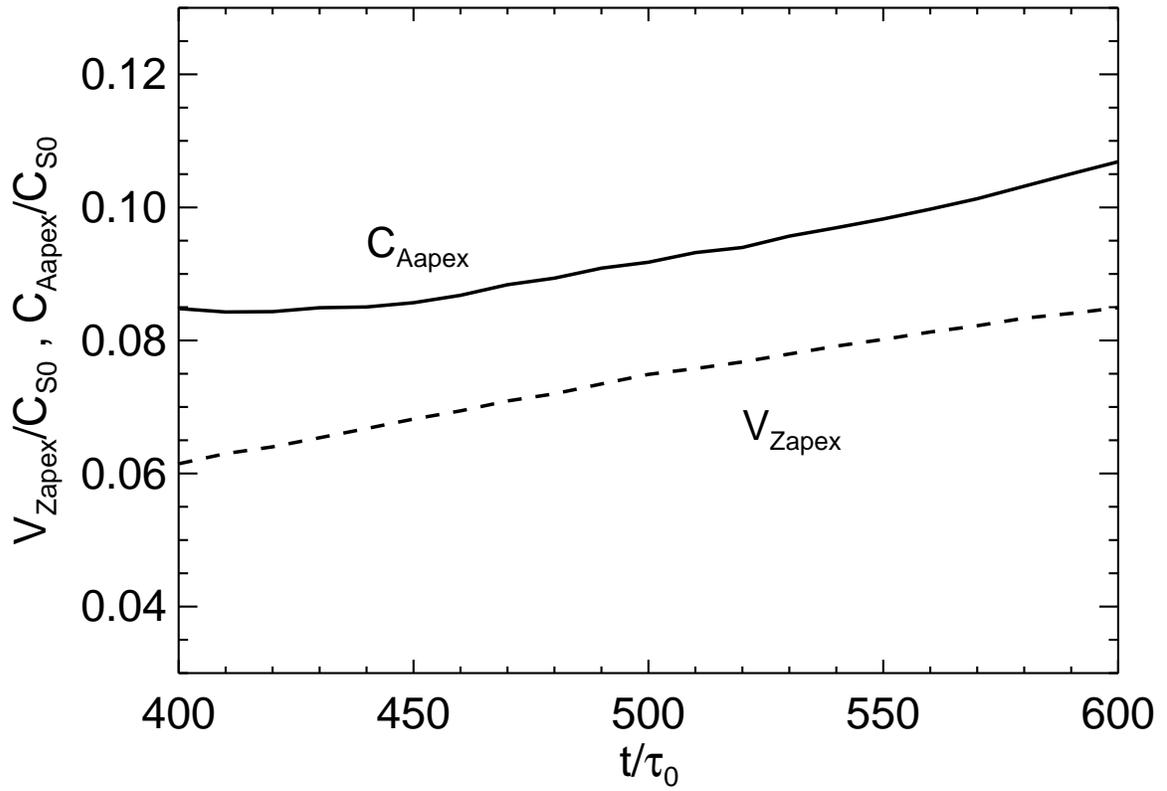}
\caption{The close-up view of Fig. \ref{fig:z_vz_top}.
Upward velocity at the apex of the loop is plotted
by a dashed line, while the local Alfv$\acute{\rm e}$n speed
is represented by a solid line.}
\label{fig:vz_va_top}
\end{figure}

The initial sinusoidal perturbation in the flux sheet triggers the Parker
instability so that the flux sheet begins to rise through the solar
convection zone by magnetic buoyancy (Fig. \ref{fig:ro-b}).
In this phase, the rise speed enhances continuously as the flux sheet
emerges.
The rising flux becomes arch-like shape owing to the stronger
buoyancy of the loop center.
The evacuation in the apex by the downflow
along the field lines due to gravity
leads to the acceleration of the loop
until reaching the local Alfv$\acute{\rm e}$n speed
($C_{\rm A}=B/\sqrt{4\pi\rho}$).
Figure \ref{fig:vz_va_top} is a close-up view of the evolution
between $t/\tau_{0} =400$ and $t/\tau_{0} =600$
of Figure \ref{fig:z_vz_top},
where a dashed line indicates the
rise velocity of the apex, while the local Alfv$\acute{\rm e}$n speed
is represented by a solid line.
This figure shows that the rise velocity increases so as to get close
to the local Alfv$\acute{\rm e}$n speed.

\subsection{Second Stage ($700<t/\tau_{0}<1900$)}
At around $t/\tau_{0}=700$, the rise velocity changes from acceleration to
deceleration (Fig. \ref{fig:z_vz_top}),
and at $t/\tau_{0}=1000$, both legs of the loop
become vertical (Fig. \ref{fig:ro-c}).
The central part of the emerging loop
flattens and expands horizontally along the surface.
Drained gas from the apex flows down along the field lines to each
trough, so that the both troughs sink into the deep convection zone.

In this stage, the rise motion turns into the deceleration
as seen from Figure \ref{fig:z_vz_top}.
\citet{maga01} found that
the tube reduces its rise speed and becomes flattened
since the rise motion cannot persist through the
convectively stable photosphere.
\citet{mur06} explained that the deceleration process occurs because the
downward pressure gradient exceeds the upward magnetic buoyancy
when the emerging flux tube is close enough to the surface.
In addition, \citet{mur06} found a period
when the rise speed diminishes due to the aerodynamic drag
exerted by the flows surrounding the tube
while in the convection zone.
In our case, however, the deceleration occurs at much deeper level
($z\sim -50H_{0}=-10,000\ {\rm km}$)
than the previous studies ($z\sim -850\ {\rm km}$).
Moreover, our simulation is carried out
in a two-dimensional scheme,
while the previous studies were done in 2.5D or 3D,
so that a three-dimensional force such as
the aerodynamic drag does not exert on our emerging loop.
Our results are explained by another mechanism as follows.

\begin{figure}
\epsscale{1.}
\plotone{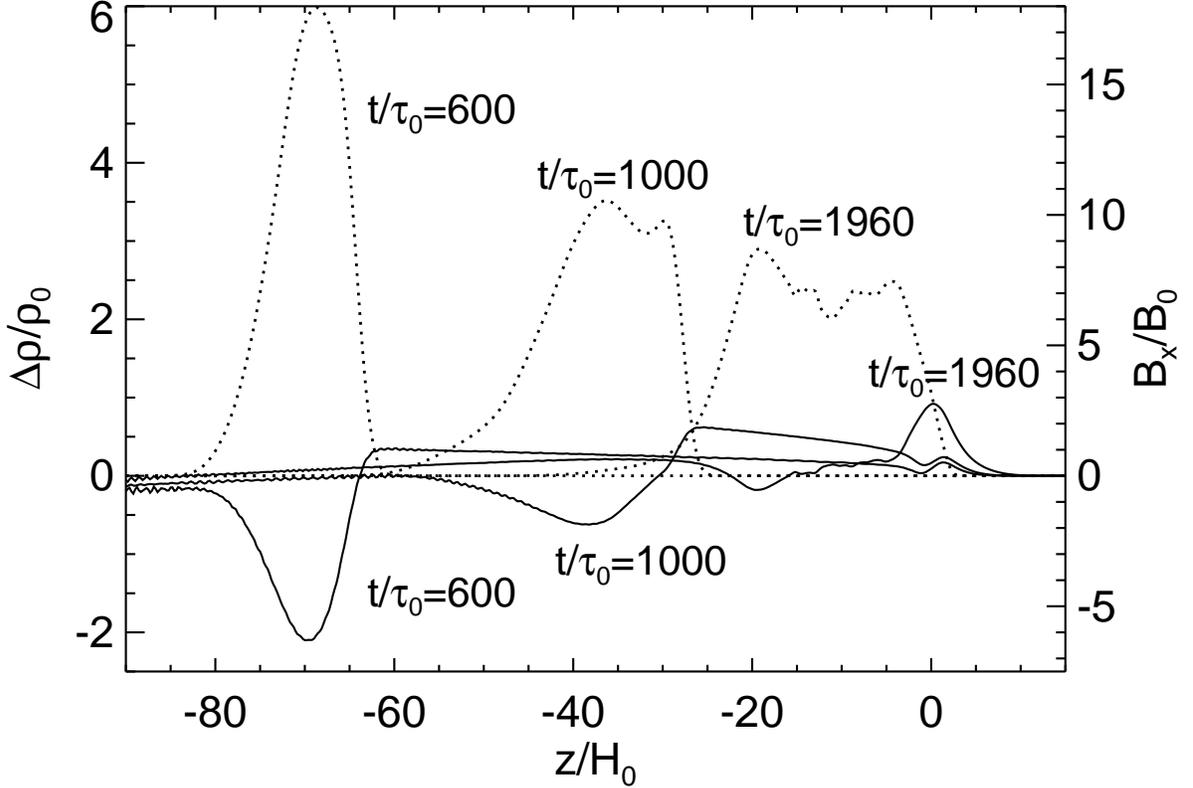}
\caption{Density differences from the initial state
($\Delta\rho\equiv\rho(t)-\rho(0)$, solid lines)
and horizontal field components ($B_{x}$, dotted lines)
along the $z$-axis at the center of the simulation box
($x/H_{0}=0$) at the three different times.
As the magnetic flux rises, mass piles up on the loop.
However, the mass cannot persist through the isothermal
photosphere ($0<z/H_{0}<10$).}
\label{fig:dro}
\end{figure}

\begin{figure}
\begin{center}
\subfigure{\includegraphics[clip,scale=0.5]{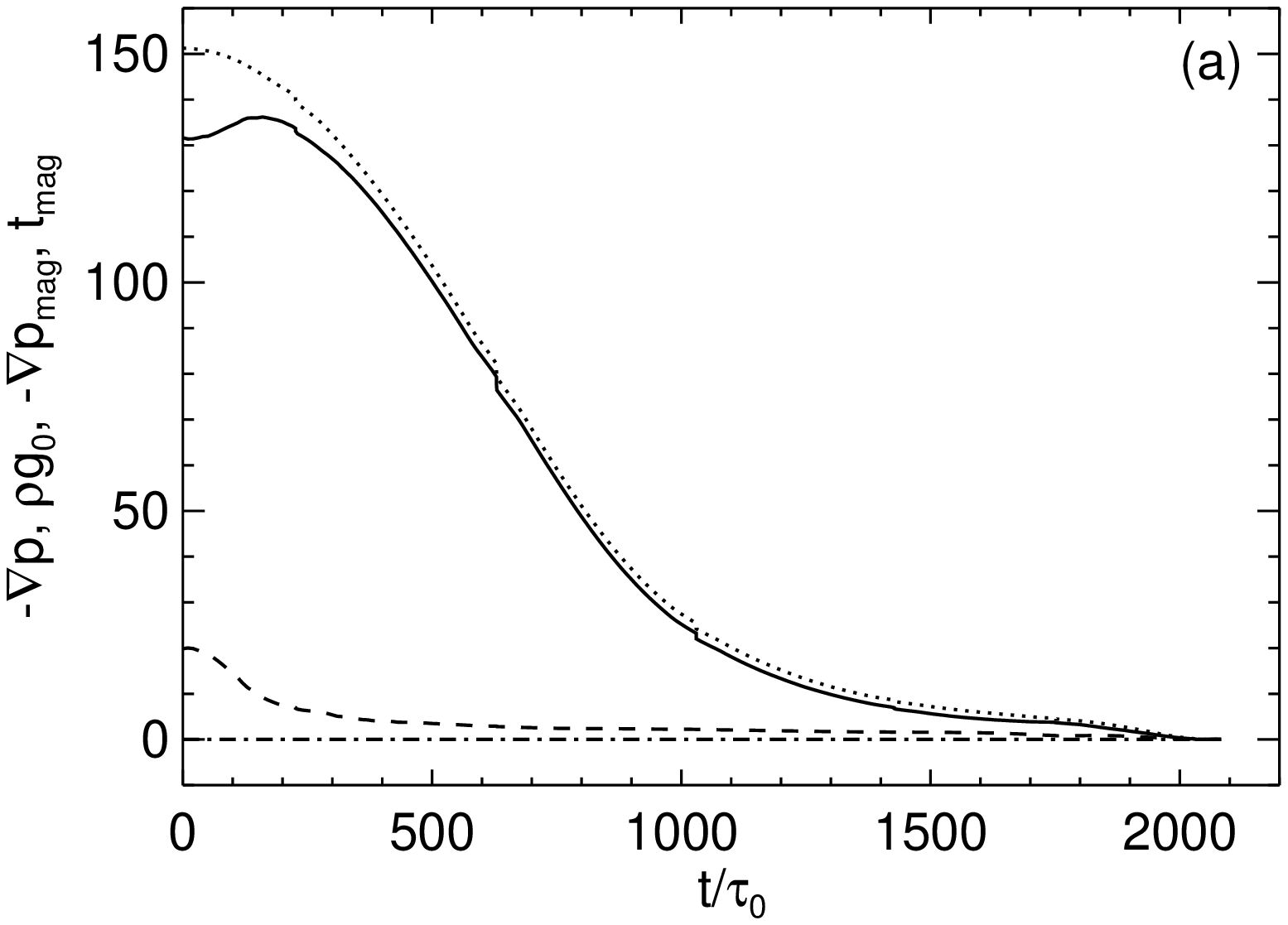}
\label{fig:force}}\\
\subfigure{\includegraphics[clip,scale=0.5]{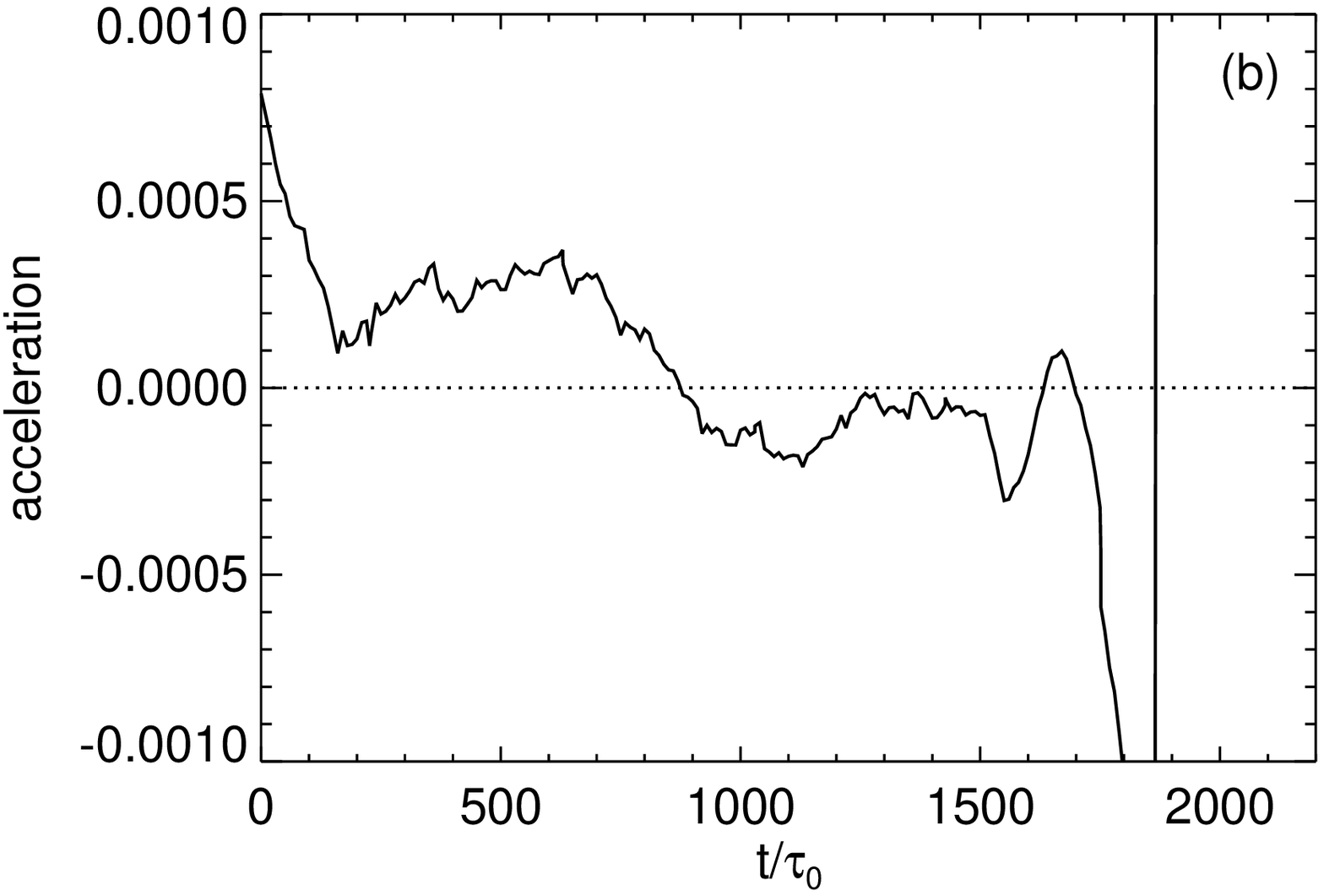}
\label{fig:accel}}
\caption{\subref{fig:force}: Time evolution of the vertical components
 of the forces acting on the apex of the emerging loop: gas pressure
 gradient ($-\nabla p=-dp/dz$, solid line), gravity ($\rho g_{0}$,
 dotted line), magnetic pressure gradient
($-\nabla p_{\rm mag}=-d[B^{2}/(8\pi)]/dz$, dashed line),
and magnetic tension
($t_{\rm mag}=[(\mbox{\boldmath
 $B$}\cdot\nabla)\mbox{\boldmath $B$}]_{z}/(4\pi)$,
dashed-dotted line).
\subref{fig:accel}: The same of the acceleration
($F_{z}/\rho=(-\nabla p-\rho g_{0}-\nabla p_{\rm mag}-t_{\rm mag})/\rho $).
A dotted line shows that the acceleration equals zero.
}
\label{fig:foracc}
\end{center}
\end{figure}

Figure \ref{fig:dro} shows the distribution
of the density difference from the initial state
($\Delta\rho\equiv\rho(t)-\rho(0)$) and the horizontal
component of the magnetic field ($B_{x}$) along the $z$-axis
at $t/\tau_{0}=600$, $t/\tau_{0}=1000$, and $t/\tau_{0}=1960$.
As can be seen in Fig. \ref{fig:dro},
the mass piled on the emerging loop cannot
rise through the photosphere/chromosphere ranging from $z/H_{0}=0$ 
to $z/H_{0}=10$ because this isothermal (i.e. strongly sub-adiabatic)
layer is convectively stable.
Figure \ref{fig:force} shows the vertical component of the forces
acting upon the apex of the loop.
They are gas pressure gradient $-\nabla p=-dp/dz$,
gravity $\rho g_{0}$,
magnetic pressure gradient
$-\nabla p_{\rm mag}=-d[B^{2}/(8\pi)]/dz$,
and magnetic tension 
$t_{\rm mag}=[(\mbox{\boldmath $B$}\cdot\nabla)
\mbox{\boldmath $B$}]_{z}/(4\pi)$.
Figure \ref{fig:accel} shows the acceleration
calculated by dividing the total force with the gas density
($F_{z}/\rho=(-\nabla p-\rho g_{0}-\nabla p_{\rm mag}-t_{\rm mag})/\rho$).
It should be noted that the total force $F_{z}$ is much smaller
than each force since the rising loop is almost in a mechanical equilibrium.
Figure \ref{fig:accel} indicates that the acceleration turns from
positive to negative at around $t/\tau_{0}=870$,
which means the rise velocity changes into deceleration phase
at that time.

The deceleration of the crest and the continuous rise motion of the both
sides cause the loop flattened, which, in turn, makes the mass left on
the flattened loop.
As a result, the rising flux decelerates and stretches horizontally
beneath and around the surface.
This process can possibly explain the formation of the `initial flux'
of the previous studies in much smaller scales
\citep[e.g.][etc.]{shiba89}.
We have to carry out three dimensional experiments
because another effects such as the aerodynamic drag would act
on the actual expanding loops.
However, the above-mentioned scheme can explain the deceleration in the
convection zone if the emerging field has a sheet-like shape
rather than a tubular form.

\subsection{Third Stage ($1900<t/\tau_{0}<2000$)\label{third}}
\begin{figure}
\epsscale{1.}
\plotone{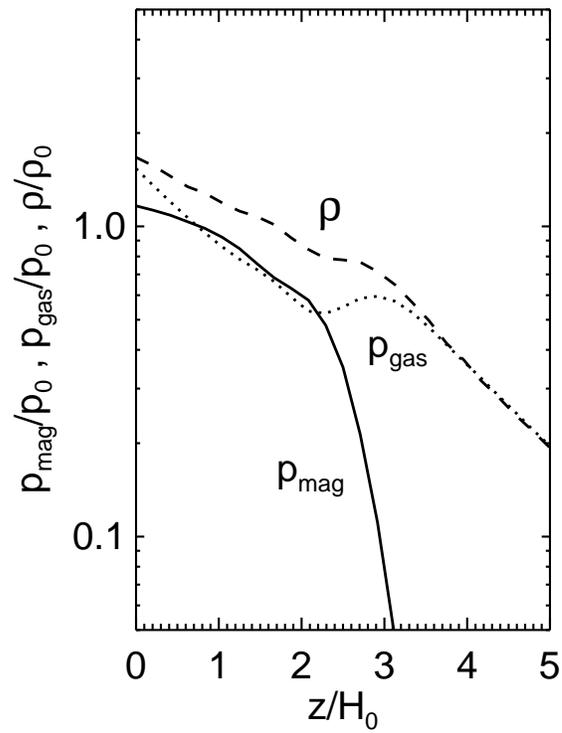}
\caption{Vertical distributions of magnetic pressure (solid line),
gas pressure (dotted line), and gas density (dashed line)
along the axis $x/H_{0}=20$ at $t/\tau_{0}=1960$,
which is the central position of the second-step emergence.}
\label{fig:surface}
\end{figure}

The field is decelerated and flattened
due to the isothermal (sub-adiabatic)
stratification on the surface;
meanwhile, the fluxes are continuously transported from beneath,
that is, the magnetic pressure gradient keeps enhancing
at the surface (Fig. \ref{fig:ro-e}).
At the location where the magnetic pressure gradient
gets steepened enough,
the further evolution to the corona occurs on the basis of the
Parker instability.
At the point of the second-step emergence,
a `pressure hill' \citep{arc04,maga01} is formed,
which indicates that the plasma in the photosphere drains out sideways
and the magnetic field covers this area, i.e., the stratification is
top-heavy (see Fig. \ref{fig:surface}).

\begin{figure}
\epsscale{1.}
\plotone{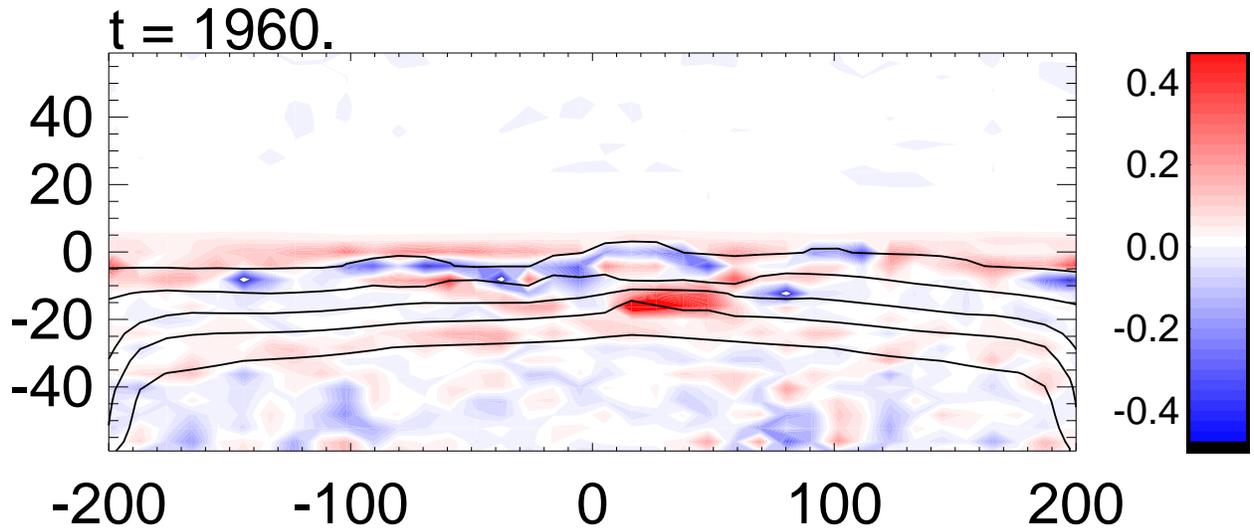}
\caption{Spatial distribution of the index
$\psi\equiv-d\rho/dz-\rho^{2}g_{0}/(\gamma p)$.
at the location shown by the white box in Fig. \ref{fig:ro-e}.
Color contour indicates $\psi$ (blue represents negative),
while magnetic field lines are overplotted by solid lines.
The aspect ratio is arranged.
The index $\psi$ is negative around
the area of the further evolution ($x/H_{0}=20,\,z/H_{0}=0$),
indicating that this area is subject to the Parker instability.}
\label{fig:newcomb}
\end{figure}

To confirm the onset of the Parker instability,
we use the criterion obtained by \citet{new61}.
The criterion for the Parker instability is \citep{new61}
\begin{eqnarray}
 -\frac{d\rho}{dz}<\frac{\rho^{2}g_{0}}{\gamma p}.
\end{eqnarray}
We plot the index
\begin{eqnarray}
 \psi\equiv -\frac{d\rho}{dz}-\frac{\rho^{2}g_{0}}{\gamma p},
\end{eqnarray}
that is, the area with negative $\psi$ is subject to the instability.
Figure \ref{fig:newcomb} illustrates the $\psi$ distribution with field lines
just before the second-step emergence at $t/\tau_{0}=1960$.
This figure indicates that the index $\psi$ is negative at around the point of
emergence; we can conclude that the further evolution to the corona
is ascribed by the Parker instability.

At $t/\tau_{0}=1960$, plasma $\beta\equiv p/(B^{2}/8\pi)$
is order of unity ($\sim 2$) and the magnetic field strength is about
700\,G at the low photosphere within the emergent area,
which are consistent with observations \citep[e.g.][]{wata08}.

\subsection{Fourth Stage ($t/\tau_{0}>2000$)}
In this final phase, the magnetic flux emerging within the photosphere
begins to expand to the solar corona by the magnetic pressure 
on the condition that the gas pressure acting on the
surface of the flux is weak enough
(Figs. \ref{fig:ro-f}-\subref{fig:ro-h}).
The expanding loop finally forms a large coronal loop.
This process is similar to that of the results of classical calculations
\citep[e.g.][etc.]{shiba89}.

\begin{figure}
\begin{center}
\subfigure{\includegraphics[clip,scale=0.4]{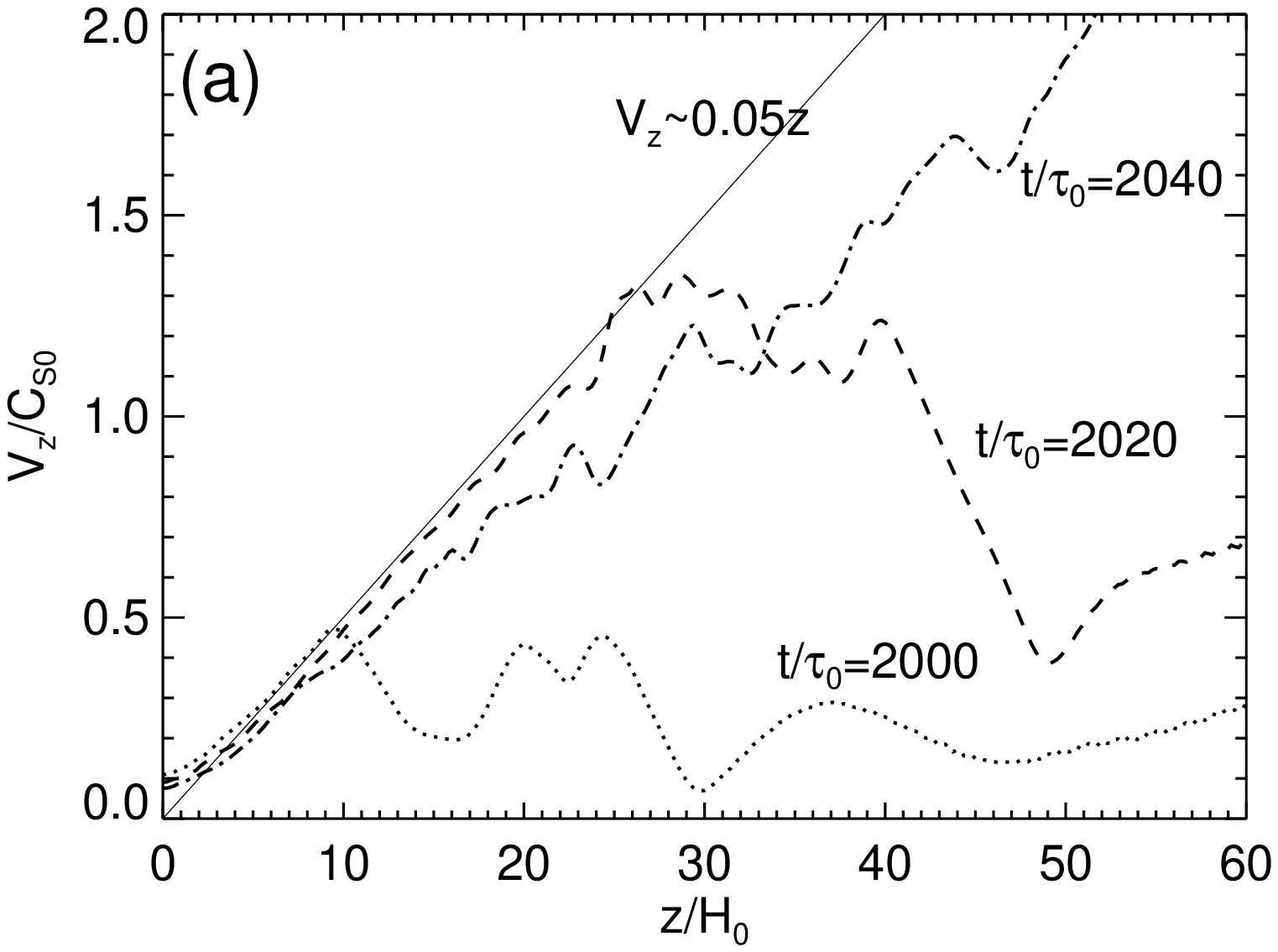}
\label{fig:nonlinear-a}}\\
\subfigure{\includegraphics[clip,scale=0.4]{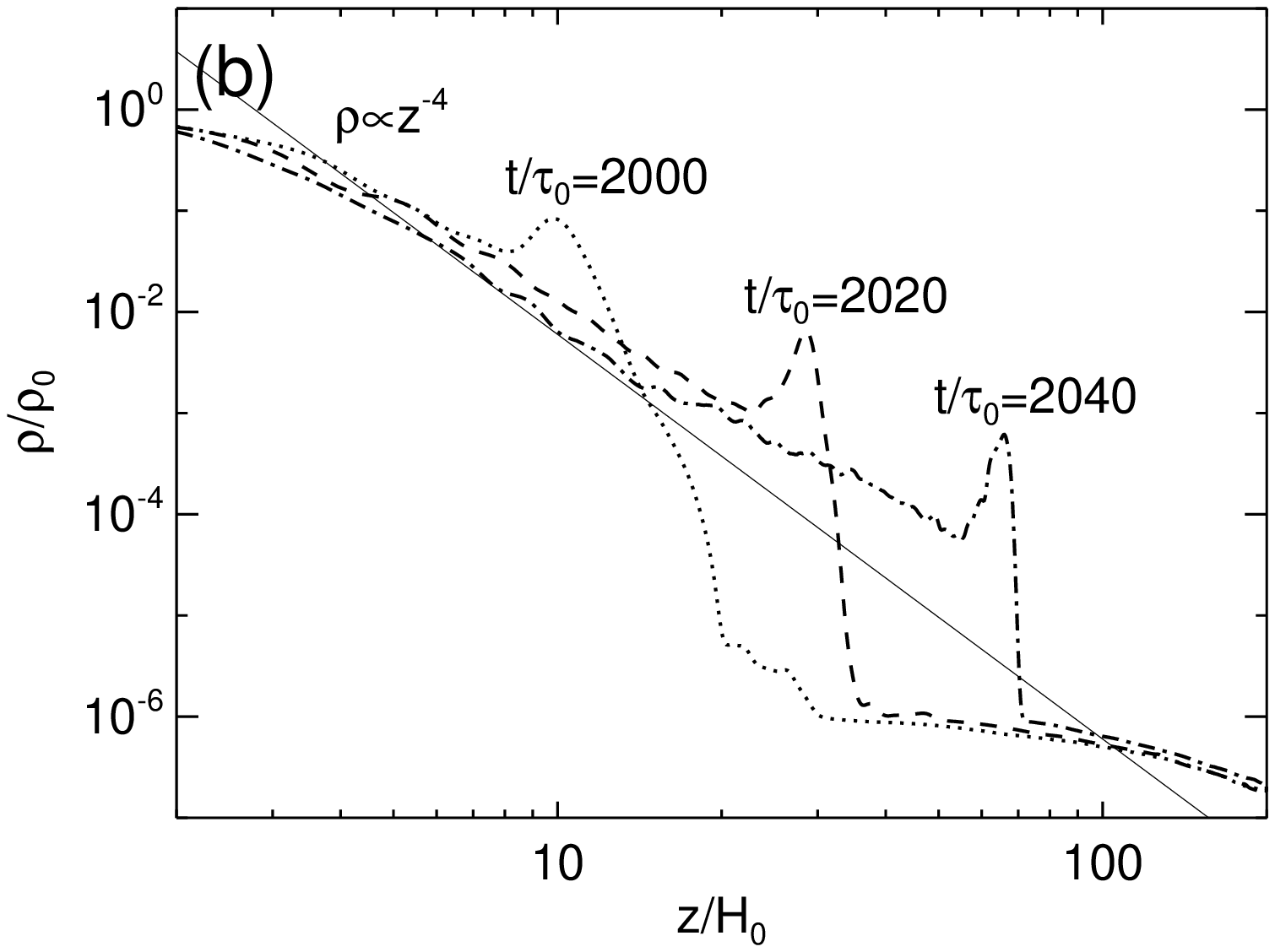}
\label{fig:nonlinear-b}}\\
\subfigure{\includegraphics[clip,scale=0.4]{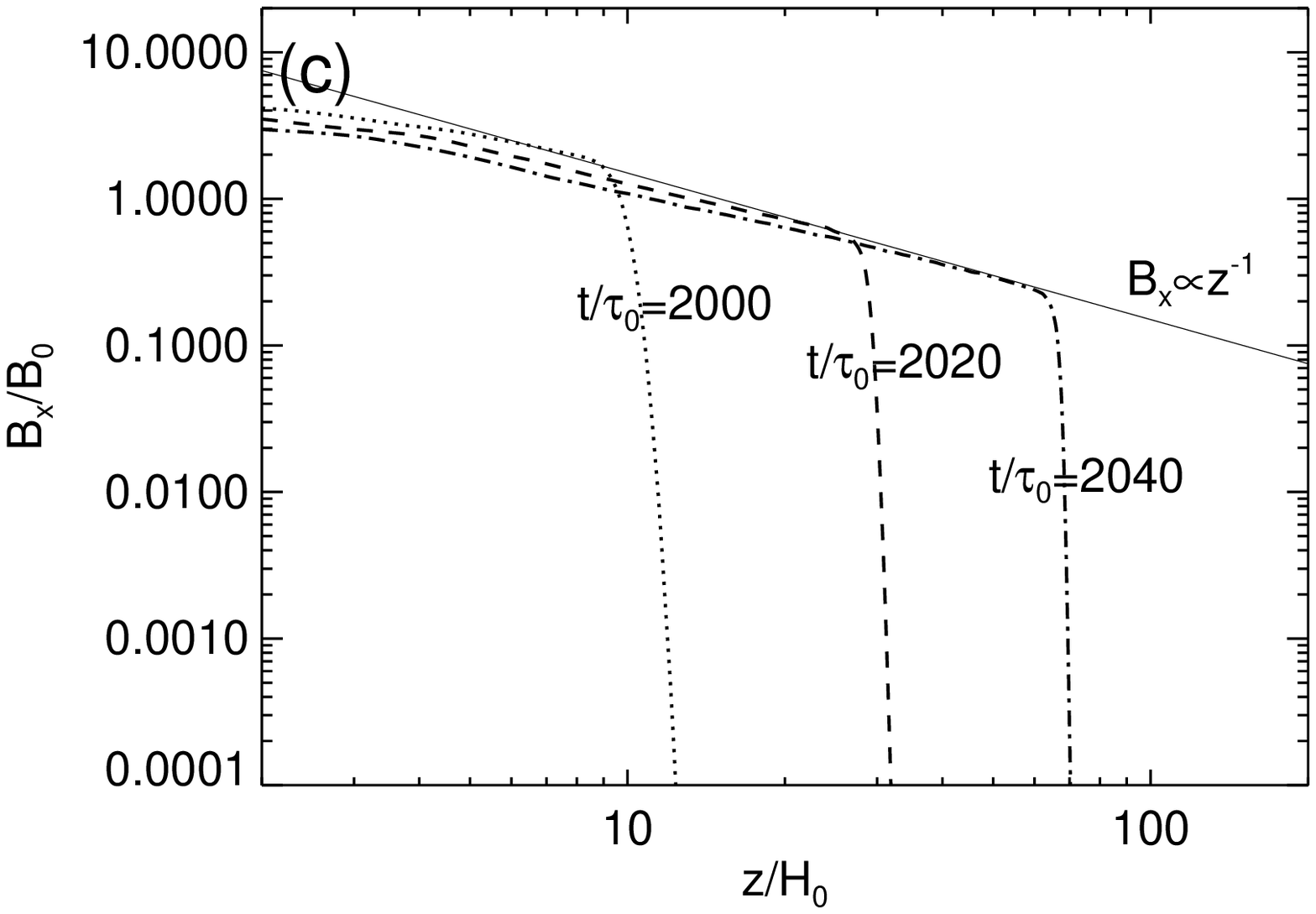}
\label{fig:nonlinear-c}}\\
\caption{\subref{fig:nonlinear-a}: Distribution of the upward velocity
along the vertical axis $x/H_{0}=20$.
Dotted, dashed, and dash-dotted lines indicate
the distribution at $t/\tau_{0}=2000$, $t/\tau_{0}=2020$,
and $t/\tau_{0}=2040$, respectively.
A solid line shows the theoretical velocity-height relation
according to \citet{shiba89}.
\subref{fig:nonlinear-b}: Distribution of the gas density
along the axis $x/H_{0}=20$.
The notation of lines is the same as in \subref{fig:nonlinear-a}.
\subref{fig:nonlinear-c}: Distribution of the horizontal component
of the magnetic field along the axis $x/H_{0}=20$.
The notation of lines is the same as in \subref{fig:nonlinear-a}.
}
\label{fig:nonlinear}
\end{center}
\end{figure}

The characteristics of this nonlinear phase is a self-similar evolution.
Figures \ref{fig:nonlinear-a}-\subref{fig:nonlinear-c}
indicate the distributions of the vertical component of
velocity, gas density, and horizontal magnetic field strength
along the axis $x/H_{0}=20$,
where the apex of the loop is located
(see Figs. \ref{fig:ro-g} and \subref{fig:ro-h}),
at $t/\tau_{0}=2000$, 2020, and 2040.
According to \citet{shiba89}, the expansion law is written as
\begin{mathletters}
\begin{eqnarray}
 V_{z}/C_{\rm s0}&=&a\,z/H_{0}, \\
 \rho&\propto&z^{-4}, \label{eq:ro}\\
 B_{x}&\propto&z^{-1}, \label{eq:bx}
\end{eqnarray}
\end{mathletters}where $a=0.05-0.07$ is about half
the non-dimensional linear growth rate
for plasma $\beta=0.5-2.0$ of the flux sheet.
In our study, plasma $\beta$ of the magnetic field
has been calculated to be 2 at $t/\tau_{0}=1960$
before the further evolution begins (see Section \ref{third}),
which suggests that the velocity-height relation is
\begin{eqnarray}
 V_{z}/C_{\rm s0}\sim 0.05\,z/H_{0}.
\end{eqnarray}
This relation is overplotted in Figure \ref{fig:nonlinear-a}
with a solid line.
The other relations given by equations (\ref{eq:ro}) and (\ref{eq:bx})
are also overplotted in
Figures \ref{fig:nonlinear-b} and \subref{fig:nonlinear-c}
with solid lines.
As seen from Figure \ref{fig:nonlinear},
the nonlinear growth to the solar corona is consistent with
that of \citet{shiba89}.

The size of the coronal loop at $t/\tau_{0}=2070$ is found to be
of $400H_{0}=80,000{\rm\,km}$ width
and $200H_{0}=40,000{\rm\,km}$ height,
while, at the surface, plasma $\beta$ is of the order of unity
and the field strength is about $4B_{0} (= 1200{\rm\,G})$.

\section{Parameter Survey and Discussion\label{parameter}}
We carry out a parameter survey by changing the values of
the initial field strength $B_{x}$ and the total magnetic flux
$\Phi$ of the emerging flux sheet to investigate the condition of the
sheet's behavior.
A summary of the values of $B_{x}$ and $\Phi$ under consideration
is given in Table \ref{tab:param}.
Figure \ref{fig:param} shows the results of the parameter survey,
where diamonds, asterisks, and X's represent `two-step,' `direct,'
and `failed' evolution, respectively.
Fluxes belong to the direct emergence group do evolve into the corona,
however, they do not reveal the deceleration,
unlike fluxes of  the two-step emergence,
while those of the failed emergence fragment
within the convection zone
or cannot pass through the photosphere.
Figure \ref{fig:z_top} indicates height-time relations
of the fluxes along the axis $x/H_{0}=0$.
In this section, we will discuss each group in detail.

\subsection{Direct emergence}
In cases 2 and 3, the fluxes show the emergence to the corona
without any deceleration at the surface.
In other words, they `directly' emerge to the corona.
As shown in Figure \ref{fig:z_top-a},
the height-time curves of cases 2 and 3 do not have
an inflection point.
We name them the `direct emergence' group.
The absence of an inflection implies that
their evolutions are not affected by the isothermal
photosphere/chromosphere at all
since they have extremely strong field
(note that these cases can be found in the upper right of
Fig. \ref{fig:param}).

Figure \ref{fig:ro_case2}
shows the time evolution of the density profile,
magnetic field lines, and velocity vectors for case 2.
Color contour is the same as that of Fig. \ref{fig:ro}.
Each flux of this group exhibits field strength $B\sim 10^{4}\ {\rm G}$
and plasma $\beta\sim 0.1$ at the surface after the emergence,
which is not consistent with observations.
Therefore, `direct emergence' model is not suited for
the formation model of active regions.

\subsection{Two-step emergence}
Cases 4-7 show the `two-step emergence' to the corona
as well as case 1 (typical model).
Height-time relations of this group are shown in
Fig. \ref{fig:z_top-b}.
Each line of this group has an inflection point beneath the surface
($z/H_{0}=0$), that is, rise velocity of the emerging flux turns from
acceleration to deceleration phase within the convection zone
due to the isothermal photosphere/chromosphere.
As the figure indicates,
the temporal length of the rising stage decreases
with increasing initial field strength and total flux.

Among these five cases, cases 4, 5, and 6 show
unrealistically strong flux densities
$B\sim 10^{4}\ {\rm G}$ (plasma $\beta\sim 0.1$) at the photosphere
after the coronal loops are built up.
The realistic models of the formation
of active regions are cases 1 and 7
($\Phi=10^{21}-10^{22}\ {\rm Mx}$ with $B_{x}=10^{4}\ {\rm G}$
at $z=-20,000\ {\rm km}$).
This result gives an important suggestion
that the fluxes which form active regions
are likely to have experienced the `two-step emergence.'

Our numerical results agree with
recent satellite data.
\citet{otsu10} found the deceleration
and the horizontal spreading
of an emerging flux within the solar chromosphere
by using {\it Hinode}/SOT.
This observation supports the concept
of the `two-step emergence' model.
%
%
At the same time, there is a difference between
our results and the observation.
The deceleration occurs in the chromosphere,
not beneath the surface.
The difference may
partly
come from the structure of the emerging loop. 
The numerical flux considered here has a sheet-like structure,
on which the plasma piles continuously during its evolution
(see Section \ref{general}).
If the emerging loop is part of a (twisted) flux tube,
the plasma on the loop can drain
around the cross-section of the tube.
Therefore, the emerging flux tube may rise through the convection zone
faster than the sheet-like flux,
and the tube may suffer a deceleration at a higher altitude
when the tube itself passes through the convectively stable layers.
On the other hand,
\citet{maga01} reported
that even in case with an initial twisted tube,
there is a deceleration close to the photosphere.
The attribution of the deceleration altitude
to the structure of the magnetic flux is
still oversimplified at this moment.
More quantitative study by three-dimensional simulations
concerning with this issue is necessary.

It is possible that the emerging flux suffers a deceleration
beneath the surface and exhibits the `two-step' evolution,
or, in some cases, fluxes show the `multi-step' emergence
since the structure of the rising loop in the convection zone
cannot be achieved by the optical observations.
To know the structure and the behavior of the flux emergence
below the photosphere,
advanced local helioseismology is needed \citep[e.g.][]{seki07}.

\subsection{Failed emergence}
\begin{figure}
\epsscale{1.}
\plotone{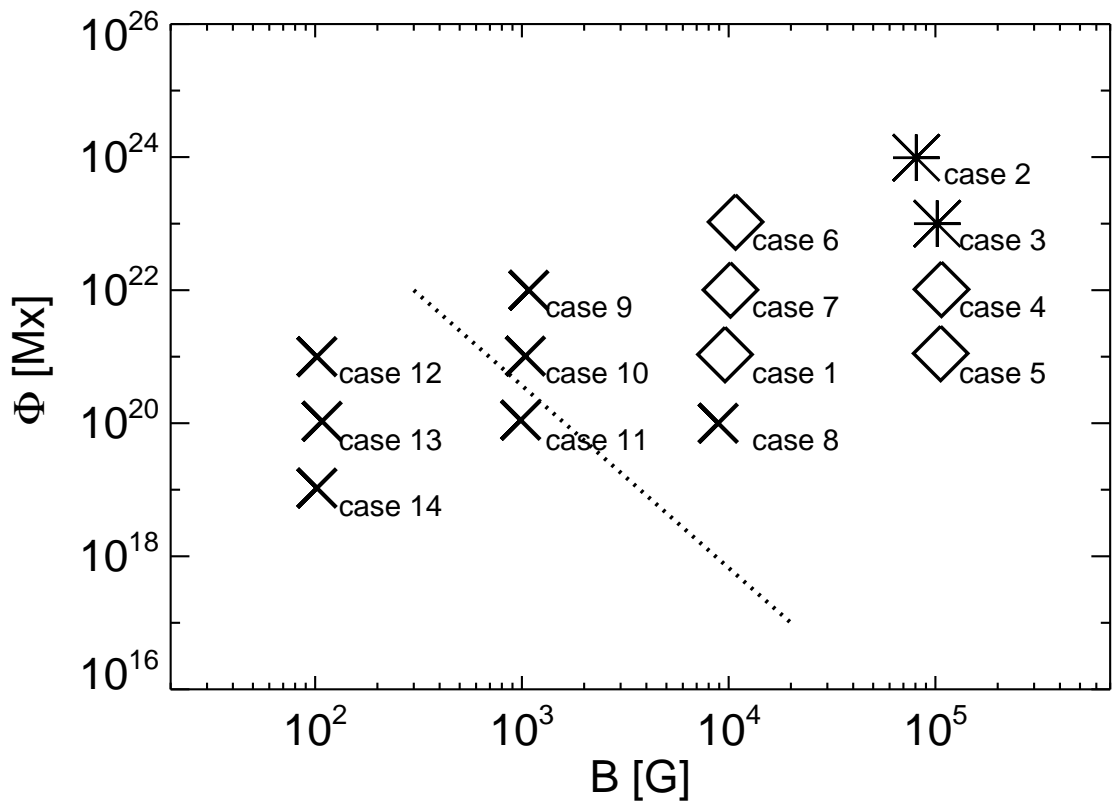}
\caption{Results of the parameter survey.
Diamonds, asterisks, and X's represent two-step emergence,
direct emergence, and failed emergence, respectively.
Case numbers are plotted on the lower right of each symbol.
Criteria for the `explosion' of the emerging flux
obtained by thin-flux-tube approximation simulations
\citep{mor95} is overplotted with a dotted line.
}
\label{fig:param}
\end{figure}

\begin{figure}
\begin{center}
\subfigure{\includegraphics[clip,scale=0.4]{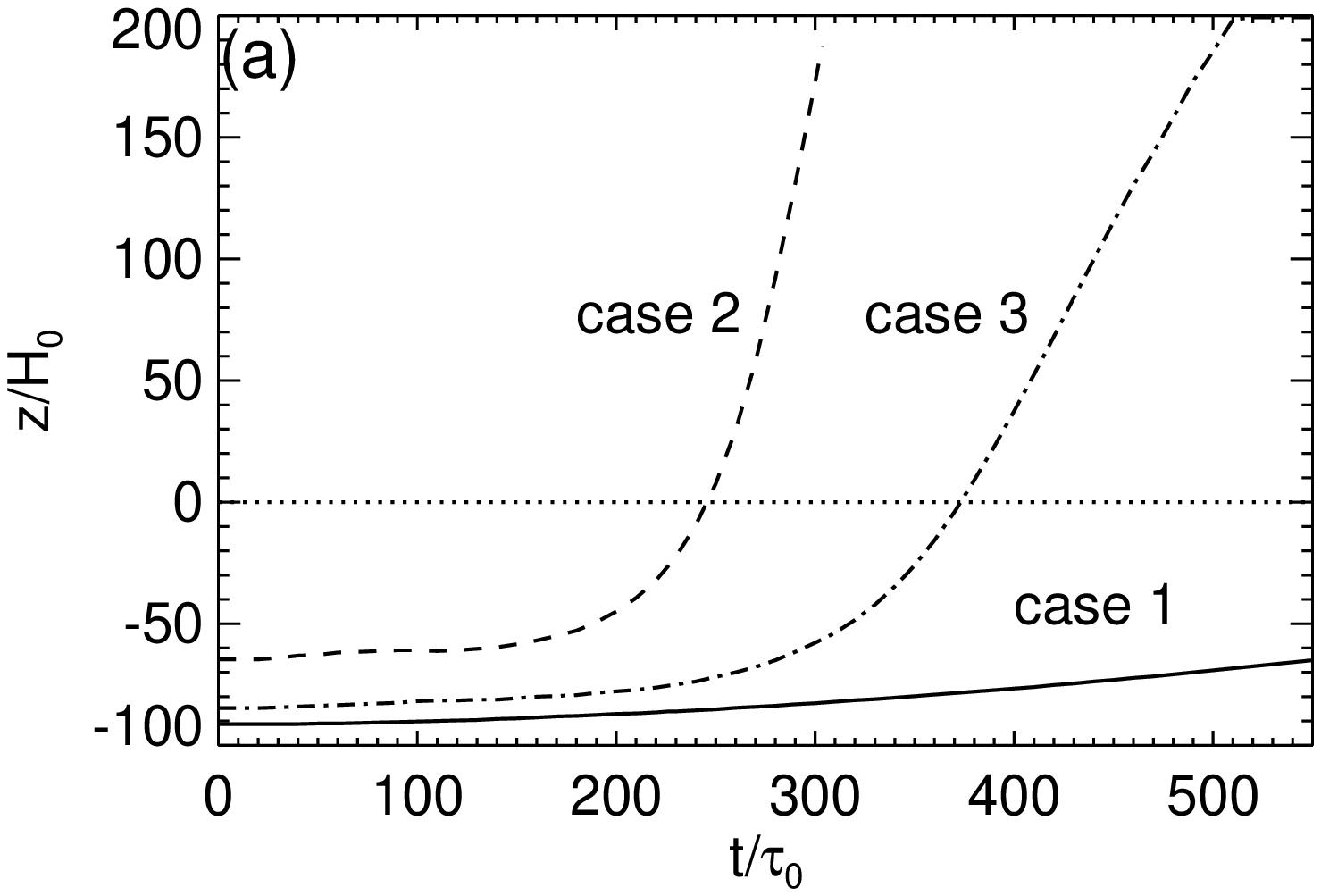}
\label{fig:z_top-a}}\\
\subfigure{\includegraphics[clip,scale=0.4]{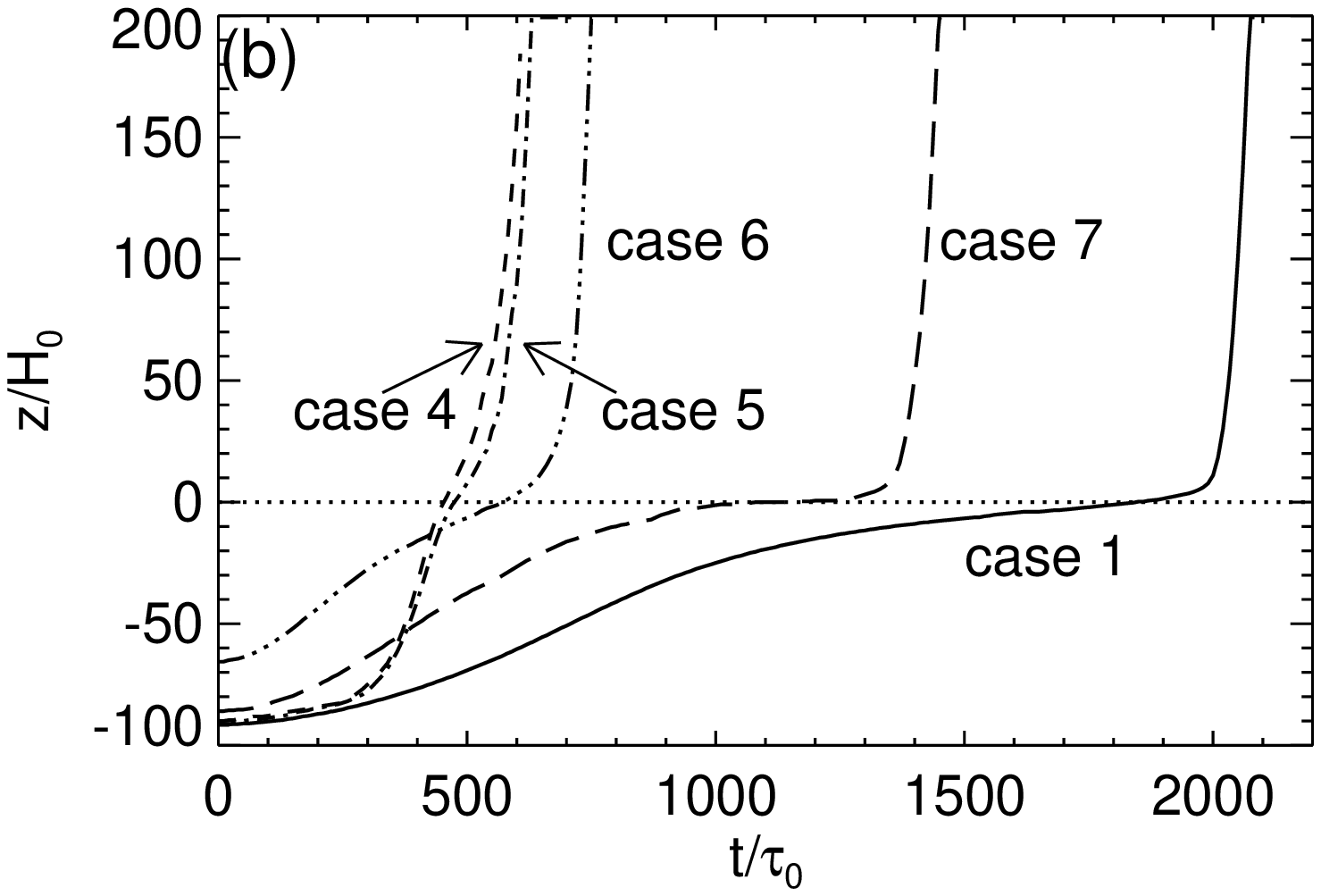}
\label{fig:z_top-b}}\\
\subfigure{\includegraphics[clip,scale=0.4]{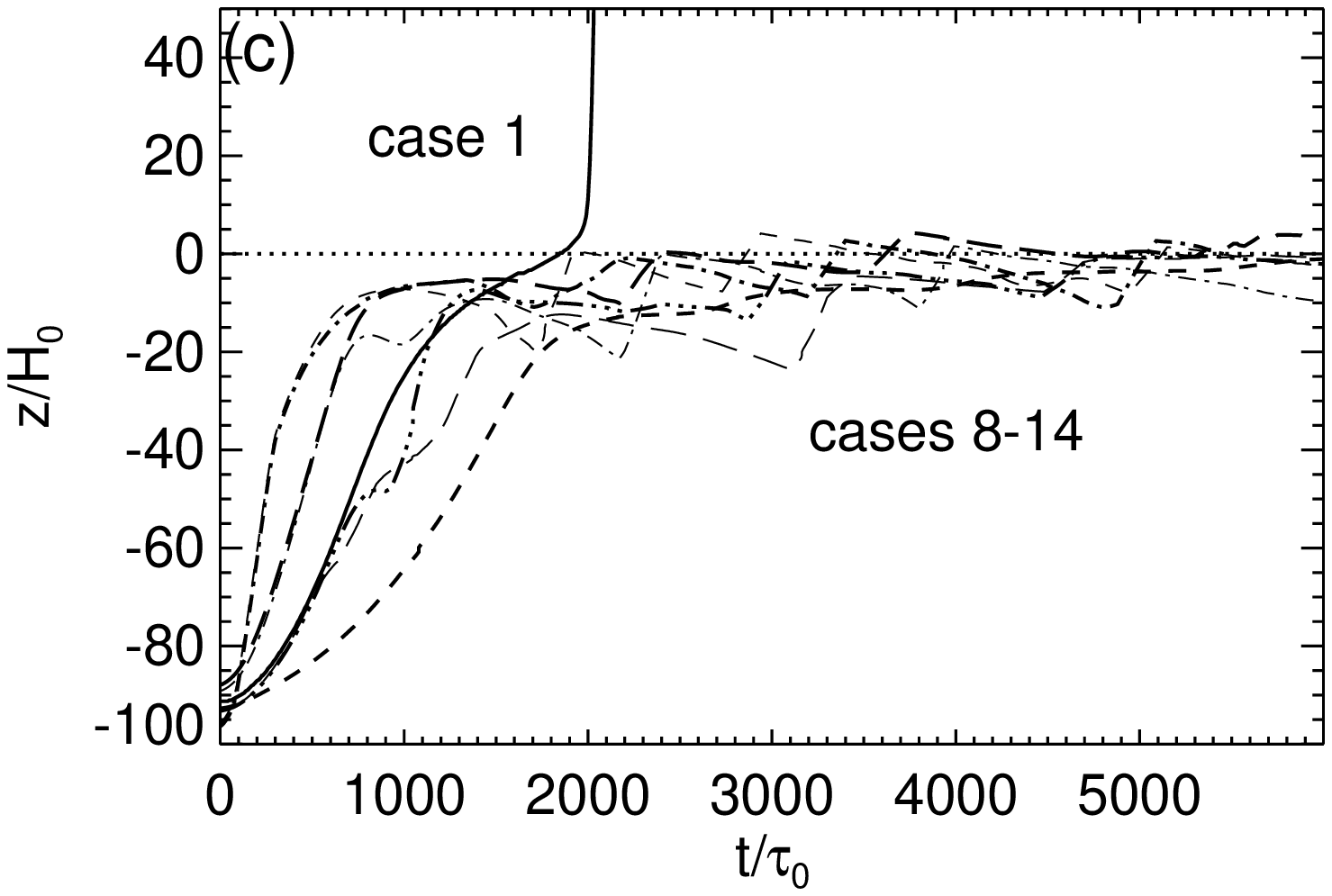}
\label{fig:z_top-c}}\\
\caption{\subref{fig:z_top-a}: Time evolution of the height of
the flux sheet along the axis $x/H_{0}=0$
for `direct emergence' (cases 2 and 3).
Dashed and dashed-dotted lines represent evolutions of cases 2 and 3,
respectively, while solid line is the height-time relation of case 1
(typical model).
\subref{fig:z_top-b}: Same for `two-step emergence' (cases 1 and 4-7).
Solid, dashed, dashed-dotted, dashed-dotted-dotted-dotted,
and long-dashed lines
represent evolutions of cases 1, 4, 5, 6, and 7, respectively.
\subref{fig:z_top-c}: Same for `failed emergence' (cases 8-14).
Dashed (thick), dash-dotted (thick), long-dashed (thick),
dashed-dotted-dotted-dotted,
dashed (thin), and dash-dotted (thin), long-dashed (thin) lines
represent evolutions of cases 8, 9, 10, 11, 12, 13, and 14,
respectively, while solid line is the height-time relation of case 1
(typical model).
In each panel, dotted line is overpoltted to show
the surface level ($z/H_{0}=0$).
}
\label{fig:z_top}
\end{center}
\end{figure}

\begin{figure}
\begin{center}
\subfigure{\includegraphics[clip,scale=0.5]{f12a.eps2}
\label{fig:ro_case2-a}}\\
\subfigure{\includegraphics[clip,scale=0.5]{f12b.eps2}
\label{fig:ro_case2-b}}\\
\subfigure{\includegraphics[clip,scale=0.5]{f12c.eps2}
\label{fig:ro_case2-c}}\\
\includegraphics[clip,scale=0.6,angle=-90.]{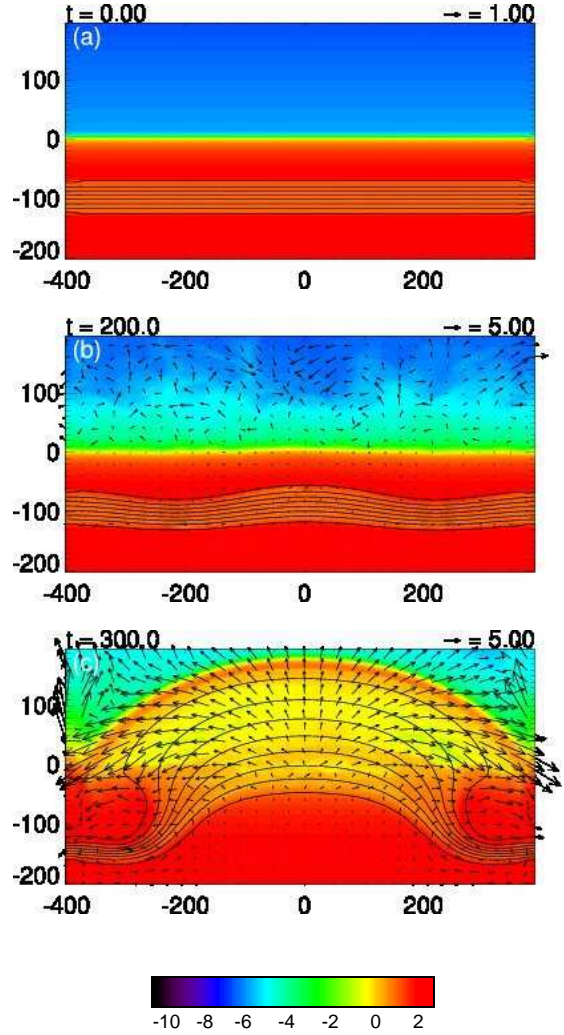}
\caption{Time-evolution of the `direct emergence' for case 2.
\subref{fig:ro_case2-a} $t/\tau_{0}=0$;
\subref{fig:ro_case2-b} $t/\tau_{0}=200$;
\subref{fig:ro_case2-c} $t/\tau_{0}=300$.
Logarithmic density profiles ($\log_{10}{(\rho/\rho_{0})}$)
are indicated by color contour,
while magnetic field lines and velocity vectors are overplotted with
black lines and arrows.}
\label{fig:ro_case2}
\end{center}
\end{figure}

\begin{figure}
\begin{center}
\subfigure{\includegraphics[clip,scale=0.5]{f13a.eps2}
\label{fig:ro_case14-a}}~
\subfigure{\includegraphics[clip,scale=0.5]{f13b.eps2}
\label{fig:ro_case14-b}}\\
\subfigure{\includegraphics[clip,scale=0.5]{f13c.eps2}
\label{fig:ro_case14-c}}~
\subfigure{\includegraphics[clip,scale=0.5]{f13d.eps2}
\label{fig:ro_case14-d}}\\
\subfigure{\includegraphics[clip,scale=0.5]{f13e.eps2}
\label{fig:ro_case14-e}}~
\subfigure{\includegraphics[clip,scale=0.5]{f13f.eps2}
\label{fig:ro_case14-f}}\\
\includegraphics[clip,scale=0.6,angle=-90.]{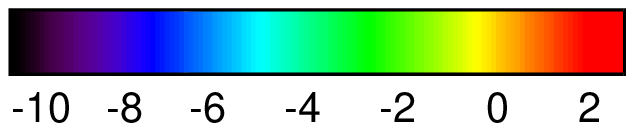}
\caption{Time-evolution of the `failed emergence' for case 14.
\subref{fig:ro_case14-a} $t/\tau_{0}=0$;
\subref{fig:ro_case14-b} $t/\tau_{0}=1000$;
\subref{fig:ro_case14-c} $t/\tau_{0}=2000$;
\subref{fig:ro_case14-d} $t/\tau_{0}=3000$;
\subref{fig:ro_case14-e} $t/\tau_{0}=4000$;
\subref{fig:ro_case14-f} $t/\tau_{0}=5000$.
Logarithmic density profiles ($\log_{10}{(\rho/\rho_{0})}$)
are indicated by color contour,
while magnetic field lines and velocity vectors are overplotted with
black lines and arrows.}
\label{fig:ro_case14}
\end{center}
\end{figure}

\begin{figure}
\epsscale{1.}
\plotone{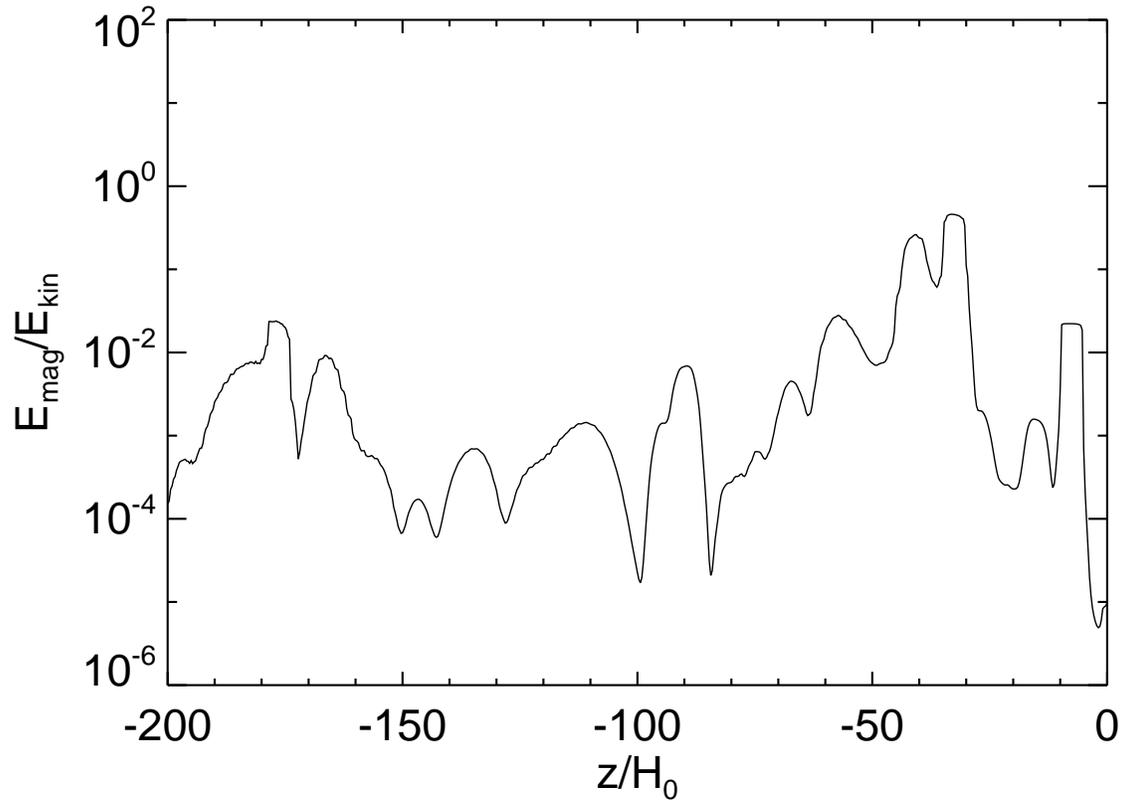}
\caption{
The ratio of $E_{\rm mag}=B^{2}/(8\pi)$
to $E_{\rm kin}=\rho v^{2}/2$
of case 14 along $x/H_{0}=0$ at $t/\tau_{0}=4000$.
}
\label{fig:beq}
\end{figure}

\begin{figure}
\begin{center}
\subfigure{\includegraphics[clip,scale=0.5]{f14a.eps2}
\label{fig:ro_case8-a}}~
\subfigure{\includegraphics[clip,scale=0.5]{f14b.eps2}
\label{fig:ro_case8-b}}\\
\subfigure{\includegraphics[clip,scale=0.5]{f14c.eps2}
\label{fig:ro_case8-c}}~
\subfigure{\includegraphics[clip,scale=0.5]{f14d.eps2}
\label{fig:ro_case8-d}}\\
\subfigure{\includegraphics[clip,scale=0.5]{f14e.eps2}
\label{fig:ro_case8-e}}~
\subfigure{\includegraphics[clip,scale=0.5]{f14f.eps2}
\label{fig:ro_case8-f}}\\
\includegraphics[clip,scale=0.6,angle=-90.]{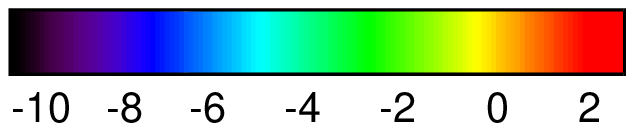}
\caption{Time-evolution of the `failed emergence' for case 8.
\subref{fig:ro_case8-a} $t/\tau_{0}=0$;
\subref{fig:ro_case8-b} $t/\tau_{0}=1000$;
\subref{fig:ro_case8-c} $t/\tau_{0}=2000$;
\subref{fig:ro_case8-d} $t/\tau_{0}=3000$;
\subref{fig:ro_case8-e} $t/\tau_{0}=4000$;
\subref{fig:ro_case8-f} $t/\tau_{0}=5000$.
Logarithmic density profiles ($\log_{10}{(\rho/\rho_{0})}$)
are indicated by color contour,
while magnetic field lines and velocity vectors are overplotted with
black lines and arrows.}
\label{fig:ro_case8}
\end{center}
\end{figure}

Cases given in Fig. \ref{fig:param} with X's belong to
the `failed emergence' group (cases 8-14).
Fluxes of cases 9-14 suffer a fragmentation
due to the continuous motion within the convection zone,
so that further emergence does not occur
(see Fig. \ref{fig:ro_case14}).
It is because the fluxes have weak fields
compared to the local kinetic energy density
of flow motion
induced by the initial perturbation.
Figure \ref{fig:beq} shows the ratio
of the magnetic energy density $E_{\rm mag}$
to the local kinetic energy density $E_{\rm kin}$
of case 14 along $x/H_{0}=0$ at $t/\tau_{0}=4000$
(Fig. \ref{fig:ro_case14-e}),
where $E_{\rm mag}=B^{2}/(8\pi)$
and $E_{\rm kin}=\rho v^{2}/2$,
respectively.
It reveals that the magnetic energy
is weaker than the kinetic energy
all over the convective layer.
Height-time relations of these fluxes
are presented in Figure \ref{fig:z_top-c}.
As the figure shows,
each line of this group reveals a continuous fluctuation
and remains around the surface,
indicating that the continuous motion repeats
within the solar interior.
On the other hand, flux of case 8
do not show the secondary emergence
because the flux fails to enhance
the magnetic pressure gradient
while crossing the surface.
This flux keeps its coherency all the while
since the field strength is almost in the same range
as the local kinetic energy density
(see Fig. \ref{fig:ro_case8}).

Dotted line in Fig. \ref{fig:param} indicates the criteria
for the `explosion' of the emerging flux obtained by
thin-flux-tube (TFT) approximation \citep{mor95}.
They found that the rising tubes with a small amount of flux
cannot reach the surface due to the `explosion':
if the tube rises sufficiently slowly,
the stratification inside the tube gets close to
a hydrostatic equilibrium along the field lines
while the stratification outside is super-adiabatic.
When the pressure difference between inside and outside the tube
is small enough at the base of the convection zone,
the difference decreases as the tube rises
because the pressure gradient inside the tube is
less steeper than that outside,
so that the magnetic field at the apex
can no longer be confined at a certain `explosion' depth.
According to their calculations,
the tubes with the initial field strength and magnetic flux
$(10^{4}\ {\rm G},\ 10^{22}\ {\rm Mx})$
and $(10^{5}\ {\rm G},\ 10^{17}\ {\rm Mx})$
at the base of the convection zone
can reach close to the surface.
Their field strength at $r/R_{\odot}=0.97$
($z=-20,000\ {\rm km}$, i.e., the depth of our initial fluxes)
are $3\times 10^{2}\ {\rm G}$ and $2\times 10^{4}\ {\rm G}$,
respectively \citep[see][Fig. 1]{mor95}.
We adopt them as criteria for the `non-explosion'.
The majority of the fluxes with parameters
in the range where they could not have reached at $z=-20,000\ {\rm km}$
according to the TFT model (left to the dotted line)
also ceases emergence even in our MHD model.

There are, however, some cases in which,
although the fluxes are expected to reach
at $z=-20,000\ {\rm km}$ level by the TFT model,
they fail to evolve further (cases 8, 9 and 10).
Especially, case 8
($10^{4}\ {\rm G}$ with $10^{20}\ {\rm Mx}$
at $z=-20,000\ {\rm km}$)
reveals the photospheric field strength
$B\sim 1\ {\rm G}$, indicating it could be the source of
the magnetic fields in the quiet sun
if the fields are enhanced,
e.g., by the flux expulsion due to the magneto-convection
at the surface.
Interestingly, this flux maintains its coherency all the while
and remains floating around beneath the surface
after it fails the further evolution to the corona
because it has a strong field that is in
equipartition with the kinetic energy density
(Fig. \ref{fig:ro_case8}).
This suggests that the flux of case 8
can be the origin of the ephemeral regions.
Flux 8 has $10^{20}\ {\rm Mx}$,
consistent with the flux of mid-sized ephemeral regions
\citep{hag01}.

\section{Conclusions\label{conclude}}
We perform the nonlinear two-dimensional simulations to investigate
the behavior of emerging flux from moderately deep convection zone
($z=-20,000\ {\rm km}$).
We set a much wider numerical box
($160\ {\rm Mm} \times 80\ {\rm Mm}$)
than those of the previous experiments
on the Parker instability \citep[e.g.][]{shiba89}.

In the typical case
($B=10^{4}\ {\rm G}$ with $\Phi=10^{21}\ {\rm Mx}$ 
at $z=-20,000\ {\rm km}$),
the results show the `two-step emergence'.
In the middle of the way of the first emergence
to the solar surface,
the flux loop turns from an acceleration phase
to deceleration
when approaching the
(sub-adiabatically stratified, i.e., convectively stable)
photosphere/chromosphere.
The emerging flux has a sheet-like shape,
thus it is difficult for the mass on the loop
to escape from the area between the loop
and the convectively stable surface.
This mass pile-up causes the loop decelerate.
The deceleration of the apex of the expanding flux
and the continuous rising of the
hillsides
make loop flattened,
which results in the plasma kept on the flux.
This deceleration mechanism is another new one
and different from those of the preceding studies
\citep{maga01,arc04,mur06}
with magnetic flux tubes in much smaller regions.
However, our result predicts the behavior
of a flux within the convection zone,
provided the flux has a sheet-like structure .
As a result of the deceleration and the flattening,
the flux spreads sideways just beneath the surface,
at which point the rise velocity of the crest of the loop is almost zero.
Meanwhile, the flux is continuously transported from below,
then the magnetic pressure gradient enhances locally in the photosphere.
We found that the further evolution to the corona occurs on the basis
of the Parker instability.
At the point of the instability, plasma $\beta$ is calculated to be
order of unity ($\sim 2$)
and the magnetic field strength is about $700\ {\rm G}$.
In the final stage, the flux shows the nonlinear evolution to the corona,
which resembles the classical experiments \citep[e.g.][etc.]{shiba89}.
The second-step evolution is described clearly by the expansion law
by \citet{shiba89}.
We find that the coronal loop exhibits $80,000\ {\rm km}$ width
with $40,000\ {\rm km}$ height,
while the field strength of each footpoint at the surface
is about $1200\ {\rm G}$.

We perform parameter runs by changing
the initial field strength $B_{x}$ and the total flux $\Phi$
to investigate the condition of the `two-step emergence'.
The results of the runs under considerations can be divided into three groups:
`direct', `two-step', and `failed' emergence.
In case of the `direct emergence',
the flux do evolve to the corona,
but they do not show the deceleration by the isothermal surface
due to their strong initial magnetic fields
($10^{23} - 10^{24}\ {\rm Mx}$ with $10^{5}\ {\rm G}$
at $z=-20,000\ {\rm km}$).
The coronal loops present irregularly strong flux densities
at the footpoints;
thus, we conclude that they are not suitable for the formation models
of active regions.
As for the cases showing the `two-step emergence',
two out of five exhibit the favorable values
of the photospheric field strength and plasma $\beta$.
The others have so large values that they cannot be regarded
as realistic models of active regions.
We can say that active regions on the sun
are likely to have undergone the deceleration and likely to show
the `two-step emergence' mentioned above.
The condition for this `two-step' active region is
ranging from $10^{21}$ to $10^{22}\ {\rm Mx}$ with $10^{4}\ {\rm G}$
at $z=-20,000\ {\rm km}$ in the convection zone.
Some recent observations support this two-step model.
The cases with $B\lesssim 10^{4}\ {\rm G}$ reveal `failed' evolutions;
they fragment within the convection zone or cannot have sufficient
magnetic pressure gradient to trigger the instability
that the second-step emergence do not occur
although the flux maintains its coherency.
We have some discussions in connection with the results
of the thin-flux-tube (TFT) model by \citet{mor95}.
The cases which are found to have `exploded' in the deeper point
in the TFT scheme also do not show further evolutions in our MHD scheme.
However, there are some cases which escape the `explosion'
fail the second-step evolutions,
one of which is possibly the source of the magnetic field
in the quiet sun.

The present calculations are in a two-dimensional scheme
solving simplified equations.
Thus we have to demonstrate the more realistic experiments in 3D.
At the same time, advanced observations
by helioseismological technique
are needed to reveal
the detail of the emerging flux in the convection zone.

\acknowledgments

Numerical computations were in part carried out on NEC SX-9
at Center for Computational Astrophysics, CfCA,
of National Astronomical Observatory of Japan.
Numerical computations were in part carried out
on Space Science Simulator (NEC SX-6)
of JAXA Supercomputer System.
Numerical computations were in part carried out on NEC SX-8
at Nobeyama Solar Radio Observatory
of National Astronomical Observatory of Japan.
The authors would like to thank H. Isobe and K. Shibata of Kyoto University,
R. Matsumoto of Chiba University,
and T. Magara of Kyung Hee University
for their fruitful suggestions and comments.




\clearpage



\clearpage

\clearpage

\clearpage

\clearpage

\clearpage

\clearpage

\clearpage

\clearpage

\clearpage

\clearpage

\clearpage

\clearpage

\clearpage

\clearpage







\clearpage

\begin{deluxetable}{crrrrr}
\tabletypesize{\footnotesize}
\tablecaption{Summary of Cases\label{tab:param}}
\tablewidth{0pt}
\tablehead{
\colhead{case} &
\colhead{$B_{x}$ [G]\tablenotemark{a}} &
\colhead{$\Phi$ [Mx]\tablenotemark{b}} &
\colhead{$\beta_{*}$\tablenotemark{c}} &
\colhead{$D$ [km]\tablenotemark{d}} &
\colhead{$N_{x}\times N_{z}$\tablenotemark{e}}
}
\startdata
 1 & $1.0\times 10^{4}$ & $1.0\times 10^{21}$ &  $1.6\times 10^{2}$ &
  1000 & $1536\times 1920$ \\
 2 & $8.1\times 10^{4}$ & $9.8\times 10^{23}$ & $1.1\times 10^{-1}$ &
11,000 & $1024\times 1280$ \\
 3 & $1.0\times 10^{5}$ & $1.0\times 10^{23}$ & $2.0\times 10^{-1}$ &
  3000 & $1024\times 1280$ \\
 4 & $1.1\times 10^{5}$ & $1.0\times 10^{22}$ & $2.0\times 10^{-1}$ &
   840 & $1024\times 1280$ \\
 5 & $1.1\times 10^{5}$ & $1.1\times 10^{21}$ & $2.0\times 10^{-1}$ &
   200 & $1024\times 1280$ \\
 6 & $1.1\times 10^{4}$ & $1.1\times 10^{23}$ &  $2.2\times 10^{2}$ &
11,400 & $1024\times 1280$ \\
 7 & $1.0\times 10^{4}$ & $1.0\times 10^{22}$ &  $1.6\times 10^{2}$ &
  3200 & $1024\times 1280$ \\
 8 & $8.9\times 10^{3}$ & $1.0\times 10^{20}$ &  $1.5\times 10^{2}$ &
   260 & $1024\times 1280$ \\
 9 & $1.1\times 10^{3}$ & $1.0\times 10^{22}$ &  $2.2\times 10^{4}$ &
11,000 & $1024\times 1280$ \\
10 & $1.0\times 10^{3}$ & $1.0\times 10^{21}$ &  $1.6\times 10^{4}$ &
  3200 & $1024\times 1280$ \\
11 & $9.9\times 10^{2}$ & $1.1\times 10^{20}$ &  $1.5\times 10^{4}$ &
  1000 & $1024\times 1280$ \\
12 & $1.0\times 10^{2}$ & $1.0\times 10^{21}$ &  $2.5\times 10^{6}$ &
11,400 & $1024\times 1280$ \\
13 & $1.1\times 10^{2}$ & $1.0\times 10^{20}$ &  $1.4\times 10^{6}$ &
  3200 & $1024\times 1280$ \\
14 & $1.0\times 10^{2}$ & $1.0\times 10^{19}$ &  $1.4\times 10^{6}$ &
   960 & $1024\times 1280$ \\
\enddata
\tablenotetext{a}{Initial field strength.}
\tablenotetext{b}{Total magnetic flux.}
\tablenotetext{c}{Plasma beta at the sheet center.}
\tablenotetext{d}{Width of the sheet.}
\tablenotetext{e}{Total grid points.}
\end{deluxetable}



\begin{thebibliography}{}
\bibitem[Abbett \& Fisher(2003)]{abb03} Abbett, W. P.,
\& Fisher, G. H. 2003, \apj, 582, 475
\bibitem[Acheson(1979)]{ach79} Acheson, D. J.  1979, \solphys, 62, 23
\bibitem[Archontis et al.(2004)]{arc04} Archontis, V.,
Moreno-Insertis, F., Galsgaard, K., Hood, A., \& O'Shea, E.
2004, \aap, 426, 1047
\bibitem[Caligari et al.(1995)]{cal95} Caligari, P.,
Moreno-Insertis, F., \& Sch$\ddot{\rm u}$ssler, M.  1995, \apj, 441, 886
\bibitem[Cheung et al.(2008)]{che08} Cheung, M. C. M.,
Sch$\ddot{\rm u}$ssler, M., Tarbel, T. D., \& Title, A. M.
2008, \apj, 687, 1373
\bibitem[D'Silva \& Choudhuri(1993)]{dsil93} D'Silva, S.,
\& Choudhuri, A. R.  1993, \aap, 272, 621
\bibitem[Fan(2001a)]{fan01a} Fan, Y.  2001a, \apj, 546, 509
\bibitem[Fan(2001b)]{fan01b} Fan, Y.  2001b, \apjl, 554, L111
\bibitem[Fan(2004)]{fan04} Fan, Y.  2004, Living Rev. Solar Phys., 1, 1
\bibitem[Gough(1969)]{gou69} Gough, D. O.  1969,
J. Atmos. Sci., 26, 448
\bibitem[Hagenaar(2001)]{hag01} Hagenaar, H. J. 2001, \apj, 555, 448
\bibitem[Lantz \& Fan(1999)]{lan99} Lantz, S. R., \& Fan, Y.
1999, \apjs, 121, 247
\bibitem[Magara(2001)]{maga01} Magara, T. 2001, \apj, 549, 608
\bibitem[Matsumoto et al.(1993)]{matsu93} Matsumoto, R.,
Tajima, T., Shibata, K., \& Kaisig, M.  1993, \apj, 414, 357
\bibitem[Moreno-Insertis et al.(1995)]{mor95} Moreno-Insertis, F.,
Caligari, P., \& Sch$\ddot{\rm u}$ssler, M.  1995, \apj, 452, 894
\bibitem[Murray et al.(2006)]{mur06} Murray, M. J.,
Hood, A. W., Moreno-Insertis, F., Galsgaard, K., \& Archontis, V.
2006, \aap, 460, 909
\bibitem[Newcomb(1961)]{new61} Newcomb, W. A.  1961, Phys. Fl., 4, 391
\bibitem[Nozawa et al.(1992)]{noza92} Nozawa, S., Shibata, K.,
Matsumoto, R., Sterling, A. C., Tajima, T., Uchida, Y.,
Ferrari, A., \& Rosner, R.  1992, \apjs, 78, 267
\bibitem[Otsuji et al.(2010)]{otsu10} Otsuji, K. et al.
2010, \pasj, in press
\bibitem[Parker(1955)]{par55} Parker, E. N.  1955, \apj, 121, 491
\bibitem[Parker(1966)]{par66} Parker, E. N.  1966, \apj, 145, 811
\bibitem[Parker(1975)]{par75} Parker, E. N.  1975, \apj, 198, 205
\bibitem[Sekii et al.(2007)]{seki07} Sekii, T., et al.
2007, \pasj, 59, S637
\bibitem[Shibata et al.(1989)]{shiba89} Shibata, K., Tajima, T.,
Steinolfson, R. S., \& Matsumoto, R.  1989, \apj, 345, 584
\bibitem[Spruit(1981)]{spr81} Spruit, H. C. 1981, \aap, 98, 155
\bibitem[Tsuneta et al.(2008)]{tsune08} Tsuneta, S. et al.
2008, \solphys, 249, 167
\bibitem[Watanabe et al.(2008)]{wata08} Watanabe, H., Kitai, R.,
Okamoto, K., Nishida, K., Kiyohara, J., Ueno, S., Hagino, M.,
Ishii, T. T., \& Shibata, K.  2008 \apj, 684, 736
\bibitem[Yokoyama \& Shibata(1995)]{yoko95} Yokoyama, T.,
\& Shibata, K.  1995, \nat, 375, 42
\bibitem[Yokoyama \& Shibata(1996)]{yoko96} Yokoyama, T.,
\& Shibata, K.  1996, \pasj, 48, 353
\end{thebibliography}
\end{document}